\documentclass[11pt]{article}
\pdfoutput = 1

\usepackage[utf8]{inputenc}
\usepackage{color,graphicx}
\usepackage{epsfig}
\usepackage{amsmath}
\usepackage{verbatim}
\usepackage{amssymb}
\usepackage{mathabx}
\usepackage[mathcal]{eucal}
\usepackage{stmaryrd}
\usepackage{empheq}
\usepackage{esint}
\usepackage{physics}
\usepackage{amsfonts}
\usepackage{cite}
\usepackage{array}
\usepackage{setspace}
\usepackage{braket}
\usepackage{float}
\usepackage{url}
\usepackage{mathtools}
\usepackage{tensor}
\usepackage{bbold}
\usepackage{slashed}
\usepackage{tikz}
\usepackage{pgfplots}
\usepackage{graphicx}
\usepackage{epstopdf}
\usepackage{subcaption}
\usepackage[labelfont=bf,font={sf}]{caption}
\usepackage{pgfplots}
\usepackage{mathrsfs}
\usepackage{fancybox}
\usepackage{eurosym}
\usepackage{tcolorbox}
\usepackage{tensor}
\usepackage[normalem]{ulem}
\allowdisplaybreaks
\usepackage{changepage}

\usetikzlibrary{arrows.meta, decorations.markings}

\usepackage[margin = 2.2cm]{geometry}
\usepackage[ragged]{footmisc}
\usepackage[bookmarks=true,bookmarksnumbered=false,
hyperindex=true,bookmarksopen=true,hyperfigures=true,
colorlinks=true,linkcolor=webblue,citecolor=webgreen,urlcolor=webblue,breaklinks]{hyperref}
\definecolor{webgreen}{rgb}{0, 0.5, 0}
\definecolor{webblue}{rgb}{0, 0, 0.5}
\definecolor{webred}{rgb}{0.5, 0, 0}
\definecolor{darkgreen}{rgb}{0,0.5,0}

\newcommand{\average}[1]{\left\langle #1 \right\rangle}

\def\ben{\begin{equation}}
\def\een{\end{equation}}

   \let\d=\delta 
   
     \let\r=v

\def\be{\begin{equation}}
\def\ee{\end{equation}}
\def\ba{\begin{array}}
\def\ea{\end{array}}

\def\dalemb#1#2{{\vbox{\hrule height .#2pt
       \hbox{\vrule width.#2pt height#1pt \kern#1pt
               \vrule width.#2pt}
       \hrule height.#2pt}}}

\newcommand{\bea}{\begin{eqnarray}}
\newcommand{\eea}{\end{eqnarray}}

\def\R{{{\mathbb{R}}}}

\let\tilde=\widetilde

\renewcommand{\d}{\mathrm{d}}
\renewcommand{\i}{\mathrm{i}}

\numberwithin{equation}{section}

\begin{document}

\thispagestyle{empty}
    ~\vspace{5mm}
\begin{adjustwidth}{-1cm}{-1cm}
\begin{center}
     {\LARGE \bf 
Time in gravitational subregions and in closed universes}
   \vspace{0.4in}

     {\bf Andreas Blommaert$^{1}$, Chang-Han Chen$^2$}
     \end{center}
    \end{adjustwidth}
\begin{center}
    \vspace{0.4in}
    {$^1$School of Natural Sciences, Institute for Advanced Study, Princeton, NJ 08540, USA\\
    $^2$Department of Physics, University of California, Berkeley, CA 94720, USA }
    \vspace{0.1in}
    
    {\tt blommaert@ias.edu, changhanc@berkeley.edu}
\end{center}

\vspace{0.4in}

\begin{abstract}
\noindent What are gauge-invariant local observables in a subregion in quantum gravity? How does one even define such a subregion non-perturbatively? We study these questions in JT gravity. One can define a subregion by specifying the value of the dilaton at the boundary of the region. We study conformal matter correlators in such a subregion. There is a gravitational constraint associated with York time evolution within the causal diamond of the subregion. This constraint can be leveraged to construct gauge-invariant observables in quantum gravity, using a crossed product construction. The extrinsic curvature of Cauchy slices acts as the physical clock. This is a simple example of how gauge-invariant observables can be obtained by dressing to features of a spacetime (or other fields), without the need for introducing an external observer. The entropy associated with this algebra of observables is not an area, or any boundary term. We show that gravitational constraints only give boundary formulas for entropy when gauging isometric diffeomorphisms. York time flow is merely a conformal isometry, not an actual isometry, and thus leads to bulk contributions to entropy. We repeat our construction for Milne-type closed Big-Bang universes, which may be of independent interest.
\end{abstract}

\pagebreak
\setcounter{page}{1}
\setcounter{tocdepth}{2}
\tableofcontents

\section{Introduction}
Time in gravity is confusing. Time translations are diffeomorphisms. As diffeomorphism are redundant in quantum gravity, naively, time translations are gauged. However, we clearly experience the passing of time. Therefore, a reasonable physical theory (describing our experiments) must include a notion of time which is not a redundancy. In practice, this is achieved by considering ``relational observables'' \cite{dewitt1967quantum}. The moral of relational observables is that it always makes sense to describe the passing of time relative to ``external'' features (for instance, the vibrations of atoms, or the value of the cosmological constant). The basic mathematical principle is that one decomposes the WDW constraint, which is the generator of (gauged) time translations in gravity, into some piece associated with a ``reference clock'', and a piece describing the evolution of everything else. This full WDW Hamiltonian is a gravitational constraint:
\begin{equation}
    H_\text{clock}+H_\text{universe}=0\,.\label{1.1con}
\end{equation}
Morally speaking, one may then consider wavefunctions of the conjugate $T_\text{clock}$ of the clock's ``energy'' $H_\text{clock}$. Acting on those wavefunctions, the constraint \eqref{1.1con} becomes a Schrodinger equation describing the evolution of the ``universe'' where the (quantized) variable $T_\text{clock}$ acts as the physical clock:
\begin{equation}
    H_\text{universe}=\i\hbar \frac{\d}{\d T_\text{clock}}.\label{1.2con}
\end{equation}
Different reasonable choices of clocks have been discussed in the literature. CLPW \cite{Chandrasekaran:2022cip} included a point particle with a quantized energy in the dS static patch as ``observer'' and used the conjugate $T_\text{observer}$ to the particle's energy as reference clock. A different such choice of clock was made by \cite{Chen:2024rpx}, who consider slow-roll inflation and quantize the inflaton (the cosmological constant), using this as the clock system. Using the cosmological constant as clock is standard practice in inflationary cosmology.\footnote{For recent morally related work see \cite{Etkin:2026bwd}.}

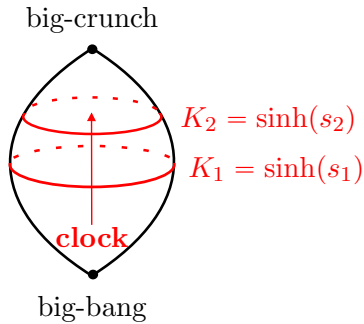
\begin{figure}
 \centering
 \begin{tikzpicture}[baseline={([yshift=-.5ex]current bounding box.center)}, scale=0.7]
 \pgftext{\includegraphics[scale=1]{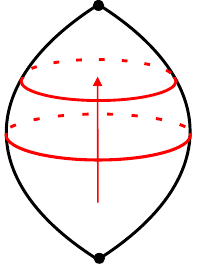}} at (0,0);
    \draw (0,2.7) node {big-crunch};
    \draw (0,-2.8) node {big-bang};
    \draw (3.35,0.8) node {\color{red}$K_2=\sinh(s_2)$};
    \draw (0,-1.43) node {\color{red}\textbf{clock}};
    \draw (3.5,-0.15) node {\color{red}$K_1=\sinh(s_1)$};
  \end{tikzpicture}
 \caption{One of the setups we study are AdS closed universes with a Milne-type big-bang and big-crunch. There is a gravitational constraint associated with conformal time flow, closely related with York time flow. In quantum gravity, where York time of the geometry becomes a dynamical (quantized) variable, we can use York time as physical clock with respect to which matter (the remainder of the universe) evolves. We dress local matter operators to this clock instead of the clock of some dynamical observer. A physical statement is: place some local  matter operator when the York time equals $K$. See \textbf{section \ref{sect:2obcl}}.}
 \label{fig:clocks}
\end{figure}

Both the constructions of \cite{Chandrasekaran:2022cip,Chen:2024rpx} are to leading order in (perturbatively small) $G$ around the classical gravitational saddle. In this work, we are instead interested in understanding how such a construction works in \textbf{non-perturbative quantum gravity} (we will, however, ignore contributions from topology change, in the spirit of addressing one problem at the time). We will study this in the simplest possible setup, namely AdS$_2$ JT gravity \cite{Maldacena:2016upp,Engelsoy:2016xyb,Jensen:2016pah}. We consider two setups, namely closed universes (with big-bangs and big-crunches of Milne type), and ``subregions'' of two-sided AdS$_2$ black holes. We study conformal matter in both setups and find explicit realizations of \eqref{1.2con}. The \textbf{clock}, which gravity chooses for us, is (roughly) the \textbf{extrinsic curvature} $K$ of Cauchy slices in these geometries. This is (related to) York time \cite{york1971gravitational,york1972role,york1973conformally,Witten:2022xxp,Parrikar:2025xmz}. We have pictured this fully dynamical (quantized) gravitational time coordinate in figure \ref{fig:clocks}. It is convenient that $K$ runs from $-\infty$ to $+\infty$ in this spacetime (unlike proper time, which has some finite range).

Both in the case of closed universes and for subregions, we obtain a type II$_\infty$ von Neumann algebra of gauge-invariant observables (for closed universes this only arises if we consider holographic conformal matter on large enough universes). Type II gravitational algebras are interesting, because they allow us to understand gravitational entropy \cite{Witten:2021unn}. In this context, we ask what is the gravitational entropy of closed universes, and of subregions. We find for closed universes equation \eqref{2.57entropyformula}:
\begin{equation}
    \delta S_\text{closed}= \frac{2\pi}{\hbar}\oint_\text{geo} \d x\sqrt{h}\,n_\mu \xi^\mu\, \delta \Phi(x)\,.\label{1.4enclosed}
\end{equation}
The integral is over a closed geodesic slice (where $K=0$) and $\xi$ is the generator of the CKV time flow. For subregions, we find a similar looking equation but now involving a non-compact spatial coordinate $x$ in equation \eqref{4.17s}:
\begin{equation}
    \delta S_\text{subregion}= \frac{2\pi}{\hbar}\int_{-\infty}^{+\infty} \d x\sqrt{h}\,n_\mu \xi^\mu\, \delta \Phi(x)\,.\label{1.4sopen}
\end{equation}
Unlike previously studied examples of \textbf{gravitational entropy}, this is a \textbf{bulk quantity}, not a boundary area (for instance). We do not understand the gravitational path integral explanation for this entropy, however we do briefly speculate on some potential interpretations. In \textbf{appendix \ref{sec:appxb}} we show that Killing vectors $\xi$ (associated with gauging isometries) result in the usual lore that gravitational constraints are boundary charges. However, non-Killing $\xi$, associated with general non-isometric diffeos, result in \emph{bulk} contributions to gravitational constraints and entropies!

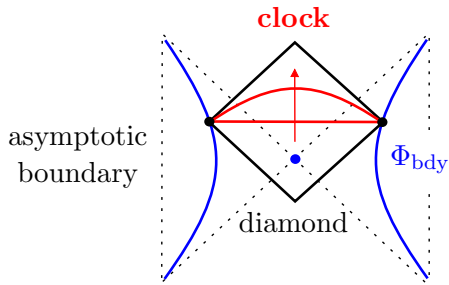
\begin{figure}
 \centering
 \begin{tikzpicture}[baseline={([yshift=-.5ex]current bounding box.center)}, scale=0.7]
 \pgftext{\includegraphics[scale=1]{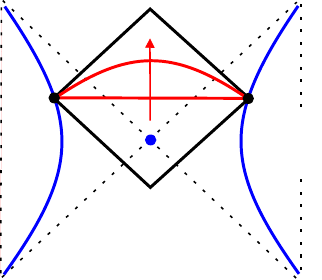}} at (0,0);
    \draw (-0.2,-1.2) node {diamond};
    \draw (-4.3,0.4) node {asymptotic};
    \draw (-4.3,-0.3) node {boundary};
    \draw (-0.2,2.7) node {\color{red}\textbf{clock}};
    \draw (2.2,0) node {\color{blue}$\Phi_\text{bdy}$};
  \end{tikzpicture}
 \caption{The second setup we consider is the causal diamond associated with a spatial subregion in global AdS. The subregion is defined by specifying the value $\Phi_\text{bdy}$ of the dilaton at the boundaries of the subregion. These are the black dots. The blue dot shows where the dilaton attains a minimum $\Phi_h$. The boost isometry may be used to place the two endpoints of the subregion at the same global time coordinate. This diamond can be sliced using curves with a fixed extrinsic curvature $K$ (or conformal time $s$). The associated generator is a constraint, allowing us to dress local matter operators to this curvature clock. The phase space consists of the geodesic volume of the subregion, the global time of the endpoints, conformal time, and its generator. See \textbf{section \ref{sect:3diamons}}.}
 \label{fig:diamondintro}
\end{figure}

One obstacle in our construction is that the definition of subregions in quantum gravity is generally poorly understood (to phrase it mildly). We provide with what seems to be a reasonable and consistent definition within JT gravity. We define a subregion by \textbf{fixing the dilaton} at the \textbf{end of the subregion}
\begin{equation}
    \Phi\rvert_\text{bdy}=\Phi_\text{bdy}\,.\label{1.5dil}
\end{equation}
We study the phase space of regions with these boundary conditions, which we can quantize. Because the dilaton plays the role of area in JT gravity, this seems a natural extension of considering subregions which ``end'' on extremal surfaces, to boundary surfaces of arbitrary (but fixed) area $\Phi_\text{bdy}$. We believe that such a definition may be used to define a version of the dS static patch in (JT) quantum gravity. In particular, whilst the boundary area is fixed, the area of the horizon (or, extremal surface) is a fully dynamical quantum variable that is allowed to fluctuate. The space of subregions can be organized by slicing causal diamonds ending on two endpoints with dilaton \eqref{1.5dil}. We picture such diamonds in figure \ref{fig:diamondintro}. One may also consider fixing the boundary dilaton gradient $p^2=g^{\mu\nu}\nabla_\mu\Phi \nabla_\nu \Phi$, as $p\rvert_\text{bdy}=p_\text{bdy}$. We briefly consider this option in the discussion \textbf{section \ref{sect:concl}}, arguing that our conclusions remain unchanged.

In additional to providing a possible avenue towards understanding observables in the static patch in quantum gravity (and the dS entropy), we have good hopes that similar techniques could be used to construct gauge invariant observables in black holes interiors, in particular in 2d dilaton gravity. This is because black hole interiors, causal diamonds (associated with subregions), and AdS closed universes share many mathematical similarities. Neither have ordinary AdS boundaries available that may serve as ``reference clocks'', and all are (in some sense) collapsing spacetimes.

The remainder of this work is organized as follows.

In \textbf{section \ref{sect:2obcl}} we construct gauge-invariant conformal matter correlators in AdS$_2$ closed universes.

In \textbf{section \ref{sect:3diamons}} we discuss the phase space and quantization of subregions in pure JT gravity (without matter).

In \textbf{section \ref{sect:4obssubr}} we construct invariant conformal matter correlators in those subregions, and discuss the associated type II$_\infty$ entropy \eqref{1.4sopen}. An important technical point allowing a straightforward construction is that the modular flow of the CFT state of the subregion is York time flow up to a reparameterization, which for ``asymptotic'' diamonds (see section \ref{sect:3.3assdiamond}) takes the form:
\begin{equation}
    K=\sinh(s)\,.
\end{equation}
In practice it is the \textbf{conformal time} $s$ of the geometry which shows up as a \textbf{gravitational clock} \eqref{1.2con} in our construction. The existence of a relation between modular flow and conformal time dates back to the seminal work of Casini-Huerta-Meyers \cite{Casini:2011kv}.\footnote{That a ``geometric'' modular flow simplifies an explicit construction of gauge-invariant local operators was, for instance, recently emphasized in \cite{Sorce:2024zme,Mertens:2025rpa}.}

In the discussion \textbf{section \ref{sect:concl}}, we select open questions that follow from this work. Examples are the (potential) applications to the dS static patch and black hole interiors; the relation of entropic formulas like \eqref{1.4enclosed} with the gravitational path integral; testing the generalized second law, etcetera.

\section{Observables in closed universes}\label{sect:2obcl}
In this section we consider closed universes in AdS JT quantum gravity. We construct gauge-invariant local observables by dressing to the extrinsic curvature $K$ of Cauchy slices. This serves as warm-up for our construction of gauge-invariant observables in subregions in section \ref{sect:4obssubr}. Closed universes in AdS JT have Milne-type big-bang and big-crunch singularities. This section may be of independent interest as perhaps the first example of constructing gauge-invariant observables in a spacetime with finite proper time. We consider full quantum gravity on a fixed topology - hence we work non-perturbatively in $G_\text{N}$\footnote{A nice feature of 2d dilaton gravity is that topology change is suppressed by a parameter $S_0$ which is independent from the Newton's constant $G_\text{N}$. This enables us to consider $S_0\to \infty$ which eliminates topology change while keeping $G_\text{N}$ finite.}. Effects from topology change are excluded in this work, but likely would be interesting and non-trivial.\footnote{See for instance \cite{Usatyuk:2024mzs,Harlow:2025pvj,Abdalla:2025gzn,Blommaert:2025bgd,Akers:2025ahe}.}

In \textbf{section \ref{sect:2.1setup}} we describe the Milne spacetimes that we will be quantizing. We furthermore discuss a thermal state for conformal matter in this geometry, which is prepared using a bra-ket wormhole \cite{Chen:2020tes}. In \textbf{section \ref{sect:2.2Ktime}} we show that the WDW Hamiltonian constraint takes the form of a Schrodinger equation for the matter theory, with extrinsic curvature $K$ acting as physical clock.\footnote{More precisely we consider $K=\sinh(s)$ where $s$ represents our chosen time coordinate on the gravitational phase space.} Because we are quantizing the full gravitational phase space, $K$ is dynamical. In \textbf{section \ref{sect:2.3closedobs}} we dress local bulk conformal matter correlators to this dynamical clock, in the same way that CLPW \cite{Chandrasekaran:2022cip} dress local operators to a dynamical observer in an otherwise (mostly) classical spacetime. In quantum gravity, when we \emph{quantize} spacetime we can dress to features of a quantized \emph{background} - like scalars \cite{Chen:2024rpx} or spacetime itself \cite{Blommaert:2025eps}. In \textbf{section \ref{sect:2.4ch}}, we investigate a special case of bulk conformal matter: a holographic 2d CFT (hence with large $c$) at a temperature above the Hawking-Page temperature. The matter operator algebra is then a type III vN algebra according to Liu-Leutheusser \cite{Leutheusser:2021frk}. The gauge-invariant algebra obtained by dressing to $K$ is an example of a crossed product construction of a type II vN algebra \cite{Witten:2021unn}. One can then rigorously study entropy differences between semiclassical states \cite{Chandrasekaran:2022cip}. This entropy is a bulk quantity (and not, for instance, some boundary area), because we consider closed universes. We will reach similar conclusions for algebras of observables in gravitational subregions in section \ref{sect:4obssubr}. The entropy is discussed in \textbf{section \ref{sect:2.4closedentropy}}.

\subsection{Setup}\label{sect:2.1setup}
\begin{figure}
 \centering
 \begin{tikzpicture}[baseline={([yshift=-.5ex]current bounding box.center)}, scale=0.7]
 \pgftext{\includegraphics[scale=1]{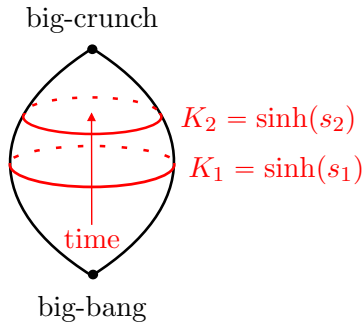}} at (0,0);
    \draw (0,2.7) node {big-crunch};
    \draw (0,-2.8) node {big-bang};
    \draw (3.35,0.8) node {\color{red}$K_2=\sinh(s_2)$};
    \draw (0,-1.45) node {\color{red}time};
    \draw (3.5,-0.15) node {\color{red}$K_1=\sinh(s_1)$};
  \end{tikzpicture}
 \caption{We investigate the classical solutions of AdS JT gravity, which are Milne-type big-bang universes with finite proper time $-\pi/2<T<\pi/2$.}
 \label{fig:milne}
\end{figure}

We consider AdS JT gravity with Euclidean action
\begin{equation}
 -I = \frac{1}{2}\int \d x \sqrt{g}\, \Phi (R+2)\label{2.1jtac}
\end{equation}
We are interested in universes without spatial boundary. The classical Lorentzian solutions are Milne-type universes, studied for instance by Louko and Sorkin \cite{Louko:1995jw}:
\begin{equation}
 \d s^2=-\d T^2+\cos(T)^2\,\d x^2\,,\quad x\sim x+B\,.
 \label{2.2metric}
\end{equation}
The total amount of proper time in this universe is finite $-\pi/2<T<\pi/2$. The beginning-and end of time are characterized by an (imaginary) scalar curvature singularity.\footnote{See for instance section 2.2 of \cite{Blommaert:2023vbz}.} We might think of these as toy models of a big-bang. The vacuum dilaton solutions takes the following form:
\begin{equation}
    \Phi=\Phi_h\sin(T)\,.
\end{equation}
The parameters $B$ and $\Phi_h$ label inequivalent classical solutions. For future reference, we note that the size $\ell$ of the closed universe and the extrinsic curvature $K$ of a fixed time $T$ Cauchy slice are given by:
\begin{equation}
    \ell=B\cos(T)\,,\quad K=\tan(T)\,.\label{2.4phasespace}
\end{equation}
In this work we will often use conformal time $s$, which is related to proper time $T$ via $\d T=\d s/\cosh(s)$ or
\begin{equation}
    \cos(T)=\frac{1}{\cosh(s)}
\end{equation}
The classical solutions become
\begin{equation}
    \d s^2=\frac{\d x^2-\d s^2}{\cosh(s)^2}\,,\quad \Phi=\Phi_h \tanh(s)\,,\quad x\sim x+B\,.\label{2.6metric}
\end{equation}
These solutions are shown in figure \ref{fig:milne}. An important equation for this work relates extrinsic curvature with conformal time:
\begin{equation}
    \boxed{K=\sinh(s)\,}\label{2.7K}
\end{equation}
Notice that $B$ and $\Phi_h$ can be gauge-invariantly defined as respectively the maximal size of the universe and the dilaton at the big crunch. These statements are mathematical precise, see \cite{Held:2024rmg} and section \ref{sect:2.2Ktime}.

We now consider conformal matter on this geometry, by coupling JT gravity to some bulk 2d CFT. Useful examples to have in mind are some collection of massless (conformally coupled) scalars, Liouville CFT, or some abstract doubly holographic 2d CFT (with large central charge $c$) \cite{Almheiri:2019hni}. We must choose a state for the CFT in order to discuss matter correlators. To maximize the analogy with our discussion of conformal correlators in subregions in section \ref{sect:4obssubr}, we choose some thermal state. To be more precise, we consider preparing the CFT matter state using a bra-ket wormhole contour \cite{Chen:2020tes}, as shown in Figure \ref{fig:braket}. In analogy with equation (6.9) in \cite{Chen:2020tes}, this assures that the CFT is prepared in a state with temperature
\begin{equation}
    \beta=\i \oint_{\mathcal{C}_{2\pi}}\d s=\i \oint_{\mathcal{C}_{2\pi}}\frac{\d T}{\cos(T)}=2\pi\label{2.8braket}
\end{equation}
Conformal matter correlators on the bra-ket wormhole contour of the spacetime \eqref{2.6metric} are then computed by conformally mapping to a Lorentzian cylinder $\d x^2-\d s^2$, $x\sim x+B$. The bra-ket wormhole contour turns this into the Euclidean torus $\d x^2+\d \sigma^2$, $x\sim x+B$, $\sigma\sim \sigma+2\pi$. A further conformal transformation turns this into the usual Euclidean torus metric $\d \varphi^2+\d \tau^2$ with $(\tau,\varphi)=(\tau+\beta,\varphi)=(\tau,\varphi+2\pi)$, where the ``inverse temperature'' experienced by conformal matter equals\footnote{We stress that conformal matter satisfies KMS \eqref{2.15kms}. The torus modulus, however, is relevant for determining the HP transition for our 2d CFT. For more details see section \ref{sect:2.4ch}.}
\begin{figure}
 \centering
 \begin{tikzpicture}[baseline={([yshift=-.5ex]current bounding box.center)}, scale=0.7]
 \pgftext{\includegraphics[scale=1]{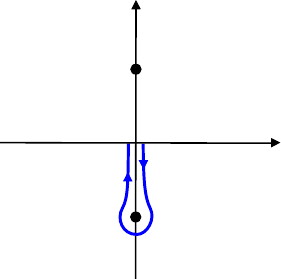}} at (0,0);
    \draw (1.5,1.2) node {big crunch};
    \draw (-0.6,1.2) node {$\frac{\pi}{2}$};
    \draw (0.45,2.15) node {T};
    \draw (-0.7,-1.75) node {\color{blue}$\mathcal{C}_{2\pi}$};
    \draw (1.45,-1.35) node {big bang};
  \end{tikzpicture}
 \caption{Bra-ket wormhole contour for the Milne spacetime preparing the CFT in a thermal state with an inverse temperature $\beta=2\pi$. One can consider alternative bra-ket wormhole contours $\mathcal{C}_{n\pi}$ \cite{Chen:2020tes}, resulting in inverse temperature $\beta=n\pi$ for conformal matter. We briefly speculate on the consequences of this freedom in the discussion section \ref{sect:concl}. Should one be summing over $n$ in the no-boundary state? Maybe not \cite{JEVwip,Held:2026huj,Blommaert:2025bgd,Aguilar-Gutierrez:2023ril}.}
 \label{fig:braket}
\end{figure}
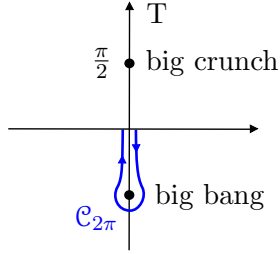
\begin{equation}\label{eq:betamat}
    \beta_\text{matter}=\frac{2\pi}{B}
\end{equation}
Correlators of conformal matter primaries $\mathcal{O}_\Delta(s)$ are then computed via the usual transformation rule:
\begin{equation}
    \average{\mathcal{O}_\Delta(s)\dots}=\cosh(s)^{\Delta} \average{\mathcal{O}_\Delta(s)\dots}_\text{torus}\,.
\end{equation}
Following Casini-Huerta-Meyers \cite{Casini:2011kv}, we henceforth consider \emph{rescaled} local operators - defined as follows:
\begin{equation}
    \tilde{\mathcal{O}}_\Delta(s)=\frac{\mathcal{O}_\Delta(s)}{\cosh(s)^\Delta}\,.\label{2.11rescale}
\end{equation}
These have the convenient property that correlators of $\tilde{\mathcal{O}}_\Delta(s)$ are exactly the 2d CFT torus correlators:
\begin{equation}
    \langle \tilde{\mathcal{O}}_\Delta(s)\dots \rangle=\average{\mathcal{O}_\Delta(s)\dots}_\text{torus}\,.\label{2.12torus}
\end{equation}
This means that correlators of $\tilde{\mathcal{O}}_\Delta(s)$ on the bra-ket contour \eqref{2.8braket} satisfy KMS. In particular, they are translation invariant under $s\to s+c$ and
\begin{equation}
    \langle \tilde{\mathcal{O}}_\Delta(s)\dots \rangle=\langle \dots\tilde{ \mathcal{O}}_\Delta(s+2\pi \i)\rangle\,.
\end{equation}
The translation invariance is an important consequence of the rescaling \eqref{2.11rescale}. Indeed, as CHM explain, KMS implies that the modular Hamiltonian of the state of the 2d CFT prepared by the bra-ket contour acts on $\tilde{\mathcal{O}}_\Delta(s)$ purely by shifting $s\to s+c$. As the modular Hamiltonian is the log of the density matrix, KMS implies the simple relation:
\begin{equation}
    \rho_\text{matter}=\rho_\text{th}=e^{-2\pi K_\text{matter}/\hbar }\,,\quad [a(s),K_\text{matter}]=\i\hbar \frac{\d}{\d s}a(s)\,.\label{2.14thermal}
\end{equation}
Here $a(s)$ denotes elements of the algebra of rescaled conformal matter operators of interest like $\tilde{\mathcal{O}}_\Delta(s)$. Equivalently one writes:
\begin{equation}
    \rho_\text{matter} a(s)=a(s+2\pi \i)\rho_\text{matter}\,.\label{2.15kms}
\end{equation}
This identification between modular flow - generated by $-\log(\rho_\text{matter})$ - and conformal time translations $s\to s+c$, allows for a straightforward construction of gauge-invariant local operators in 2d CFT coupled to AdS JT quantum gravity in the remainder of this section. That a ``geometric'' modular flow simplifies an explicit construction of gauge-invariant local operators was for instance emphasized in \cite{Sorce:2024zme,Mertens:2025rpa}. This is arguably not the physically most interesting scenario. Indeed, matter correlators which satisfy KMS are time translation invariant, and therefore cannot probe physical changes in a background as function of (modular) time. It would be of great interest to understand how to proceed without KMS. However, this goes beyond our current scope.\footnote{See also comments in \cite{Blommaert:2025bgd}.} We briefly return to these comments in the discussion (section \ref{sect:concl}).

Before proceeding me make a few more quick comments. Firstly, conformal time flow is a CKV flow with conformal Killing vector field
\begin{equation}
    \xi=\frac{\d}{\d s}
\end{equation}
Secondly, we re-emphasize that correlators of our operators of interest take exactly the form of 2d CFT Euclidean torus correlators. The modular parameter of the torus is $\tau=\i \beta_\text{matter}/2\pi=\i/B$. For instance the two point correlation function becomes the torus two-point function:
\begin{equation}
    \langle\tilde{\mathcal{O}}_\Delta(s_1)\tilde{\mathcal{O}}_\Delta(s_2)\rangle=G_{\text{torus}\,\Delta}(s_1,s_2\rvert \tau)\,.\label{2.17G}
\end{equation}
This has a non-trivial expansion into torus two-point conformal blocks. For a non-interacting CFT (the tensor product of massless scalars) this simplifies to the (double) sum of images of the Euclidean plane two-point function.\footnote{An example of an interacting CFT is Liouville theory. The equations of motion are not a linear DE due to the exponential term, therefore the method of imagines is invalidated (because a sum of solutions does not make a new solution). Gemini was useful for refreshing this fact.}

\subsection{Curvature clocks}\label{sect:2.2Ktime}
We discussed in the previous section the space of closed universes in AdS JT gravity. Next, we consider briefly its quantization following for instance \cite{Held:2024rmg}. We will first ignore matter contributions. In gravity, coordinate transformations are redundancies. This implies various constraints on the naive gravitational Hilbert space. Let's consider an example. Writing the generic 2d metric in ADM variables:
\begin{equation}\label{eq:JT_ADM}
    \d s^2 = -N^2 \d t^2 + \ell^2 (\d y+N_\perp \d t)^2\,,\quad y\sim y+1\,,
\end{equation}
and writing the JT action in first order variables, one observes that $N$ appears as a Lagrange multiplier. See for instance \cite{Iliesiu:2020zld}. The first order variables (or phase space) are $(\ell,\Phi)$ and their Legendre conjugates $(p,k)$. We work in minisuperspace, which for 2d dilaton gravity is exact due to the momentum constraints. One derives that \cite{Iliesiu:2020zld,Held:2024rmg} $k=\ell K$ with $K$ extrinsic curvature of the Cauchy slice. In quantum mechanics $[\ell,p]=\i\hbar$ and $[\Phi,k]=\i\hbar$. Then one can choose wavefunction polarizations such as $\psi(\ell,k)$, on which the Legendre conjugate variables are represented as
\begin{equation}
    p=-\i\hbar \frac{\d}{\d \ell}\,,\quad \Phi=\i\hbar \frac{\d}{\d k}\,.
\end{equation}
Path integrating over the Lagrange multiplier $N$ imposes a delta function constraint, which translates into an operator constraint:
\begin{equation}
    H_\text{grav}=-p k -\ell \Phi=0\,.\label{2.20hgrav}
\end{equation}

To solve the constraint, it is convenient to do a canonical transformation $(\ell,k)\to(S,B)$ with
\begin{equation}
    \ell=\frac{B}{\cosh(S)}\,,\quad k=B\tanh(S)\,.\label{2.21sdef}
\end{equation}
One then studies wavefunctions $\psi(S,B)$. The gravitational constraint \eqref{2.20hgrav} acts on those wavefunctions as:
\begin{equation}
    H_\text{grav}=\i \hbar \cosh(S)\frac{\d}{\d S}=0\,.\label{2.22con}
\end{equation}
This means physical gravitational wavefunctions are independent of $S$. Introducing the conjugate $K_\text{grav}$ of $S$ as a rescaled version of $H_\text{grav}$
\begin{equation}
    K_\text{grav}=\i\hbar \frac{\d}{\d S}=\frac{H_\text{grav}}{\cosh(S)}\,,\label{2.23sconj}
\end{equation}
the symplectic form becomes\footnote{Explicitly
\begin{equation}
    B=\sqrt{\ell^2+k^2}\,,\quad P=\frac{\ell p-k\Phi}{\sqrt{\ell^2+k^2}}\,.\label{2.24bp}
\end{equation}
Without matter the physical configurations satisfy $H_\text{grav}=0$. On this configuration space \eqref{2.6metric} one finds $P=-\sqrt{p^2+\Phi^2}$. This implies the usual statement $[\Phi_h,B]=\i\hbar$ discussed for instance in equation (2.12) in \cite{Held:2024rmg} (and for sine dilaton in \cite{Blommaert:2025avl}).
}
\begin{equation}
    \boxed{\d \ell\wedge \d p+\d \Phi\wedge \d k=\d K_\text{grav}\wedge \d S+\d B\wedge \d P\,}\label{2.25poisson}
\end{equation}
Imposing $K_\text{grav}=0$ means that $S$ no longer appears in the symplectic form. Therefore it is a redundant variable. In terms of states indeed we have the equivalence relation $\ket{S}\sim\ket{S+c}$ because inner products with $\bra{\psi}$ give identical answers for physical states - for which indeed $\bra{\psi}K_\text{grav}=0$. A key point is that the phase space variable $S$ represents the conformal time of the Cauchy slice of the geometry as we see in equation \eqref{2.6metric}. The previous discussion is but an elaborate way of saying that changes in conformal time are diffeomorphisms and therefore redundant in gravity, or equivalently that the there is a gravitational constraint associated with the CKV (indeed the generator of changes in conformal time). The fact that there is a constraint associated with a constraint was explained in great detail for instance in \cite{Jensen:2023yxy}. See also appendix \ref{sec:appxb}.

Note as a side comment that according to equation \eqref{2.21sdef} $B=\sqrt{k^2+\ell^2}$ indeed commutes with the constraint \eqref{2.20hgrav} and is thus indeed gauge-invariant. The same holds true for $\Phi_h=\sqrt{p^2+\Phi^2}$. See \cite{Held:2024rmg}.

Next, we include matter. The gravitational constraint is a time component of the Einstein equations. In JT gravity, the Einstein equations are second-order PDEs for dilaton:
\begin{equation}
    \nabla_\mu\nabla_\nu \Phi-g_{\mu\nu}\Box \Phi+g_{\mu\nu}\Phi=-T_{\mu\nu}\,.\label{2.26eeq}
\end{equation}
Projecting on the generator of proper time $T$ translations using the metric \eqref{2.6metric} one arrives at:
\begin{equation}
    \bigg\{\cosh(s)\frac{\d^2}{\d x^2}+\sinh(s)\frac{\d}{\d s}-\frac{1}{\cosh(s)} \bigg\}\Phi=-\sqrt{h}n^\mu n^\nu T_{\mu \nu}\,.\label{2.27Eeq}
\end{equation}
Transforming to $T$ coordinates and working in minisuperspace this becomes:
\begin{equation}
    \sin(T)\dot{\Phi}-\cos(T)\Phi=-\sqrt{h}n^\mu n^\nu T_{\mu \nu}\,.
\end{equation}
Integrating over the spatial coordinates one indeed recovers the gravitational constraint \eqref{2.20hgrav}:
\begin{equation}
    H_\text{grav}=-kp-\ell\Phi=-H_\text{matter}\label{2.28wdw}
\end{equation}
Here we used $p=-\dot{\Phi}$. To construction gauge-invariant local observables, we are interested (following for instance \cite{Sorce:2024zme,Mertens:2025rpa}) in leveraging the constraint associated with the CKV flow. The associated generator is
\begin{equation}
    K_\text{matter}=\oint \d x\,\sqrt{h}n^\mu \xi^\nu T_{\mu \nu}=\frac{H_\text{matter}}{\cosh(s)}\,.\label{2.30kmat}
\end{equation}
Recalling the definition \eqref{2.23sconj} of the canonical conjugate $K_\text{grav}$ of the dynamical gravitational variable $s$ and the constraint \eqref{2.28wdw} one obtains the following gravitational constraint:
\begin{equation}
    \boxed{K_\text{grav}+K_\text{matter}=0\,}\label{2.30constraint}
\end{equation}

Equation \eqref{2.30constraint} will be used in section \ref{sect:2.3closedobs} to construct a set of gauge-invariant local observables in a manner that parallels the CLPW construction of observables in the dS static patch \cite{Chandrasekaran:2022cip}. The analogue of the constraint \eqref{2.30constraint} in CLPW is
\begin{equation}
    q+H_\text{matter}=0\,,
\end{equation}
where $q$ is the quantized energy of an observer included in the static patch. The key difference between our construction and CLPW is that in CLPW the background geometry is semiclassical, and an observer is added; whereas in our case the background is fully quantized and no additional observer is included. Indeed, we strongly emphasize that $K_\text{grav}$ is a (dynamical) \emph{gravitational} phase space variable. We dress to features of \emph{spacetime}. In this sense, our construction is more reminiscent of that of \cite{Chen:2024rpx}.

Before proceeding, we note that what we will be doing can be phrased as using ``York time'' \cite{york1971gravitational,york1972role,york1973conformally,Witten:2022xxp}. York time in the context of JT gravity was studied in \cite{Parrikar:2025xmz} (which was partial motivation for this work). York time and conformal time (in this section) are related by the reparameterization $K=\sinh(s)$. For subregions in section \ref{sect:3diamons} this relation becomes slightly more complicated, but they are still related by a reparameterization.\footnote{One difference between our analysis and \cite{Parrikar:2025xmz} is that equations simplify when writing phase space in terms of $S$ (or $K$) and a gauge-invariant length variable $B$, rather than using $\ell$ and $K$. This is the technical reason why the constraint \eqref{2.22con} simplifies as compared to the key equations in \cite{Parrikar:2025xmz}.}

\subsection{Relational observables}\label{sect:2.3closedobs}
We now construct gauge-invariant operators. By definition, gauge invariant operators are operators that commute with the constraint \eqref{2.30constraint}, such that they map physical states (annihilated by the constraint) to physical states. The gauge-invariant operators in our theory consist, in part, of the algebra discussed in section 2 of \cite{Witten:2023xze}. Firstly, there are function of $K_\text{grav}$:
\begin{equation}
    f(\mathbf{K}_\text{grav})\,.\label{2.33}
\end{equation}
Oftentimes we use bold notation to distinguish operators (such as $\mathbf{x}$) from classical phase space variables or eigenvalues (such as $x$). Secondly, there are dressed local matter correlators, where the gravitational variable $S$ appears as the time coordinate:
\begin{equation}
    \boxed{a(s+\mathbf{S})=e^{\i K_\text{matter}\mathbf{S}/\hbar}\,a(s)\,e^{-\i K_\text{matter}\mathbf{S}/\hbar}\,}\label{2.34operators}
\end{equation}
There gauge-invariance is checked for instance through the following elementary calculation. Note first that matrix elements of these operators between $K_\text{grav}$ eigenstates are time-smeared matter observables
\begin{equation}
\bra{K_1}a(s+\mathbf{S})\ket{K_2}=\int_{-\infty}^{+\infty} \d S\, e^{\i S(K_1-K_2)/\hbar}\,a(s+S)\,.
\end{equation}
Using integration by parts one then indeed checks gauge-invariance:
\begin{align}
    \bra{K_1}[K_\text{matter},a(s+\mathbf{S})]\ket{K_2}&=-\i \hbar \int_{-\infty}^{+\infty} \d S\, e^{\i S(K_1-K_2)/\hbar}\,\frac{\d}{\d s}a(s+S)\nonumber\\
    &=-(K_1-K_2)\int_{-\infty}^{+\infty} \d S\, e^{\i S(K_1-K_2)/\hbar}\,a(s+S)\nonumber\\
    &=-\bra{K_1}[\textbf{K}_\text{grav},a(s+\mathbf{S})]\ket{K_2}\,.
\end{align}
Here we assumed that contributions to matter correlators from $s\to\pm\infty$ vanish, which is indeed correct for the thermal matter state \eqref{2.14thermal} which we consider. Thirdly, one finds the algebra of polynomials of $\mathbf{B}$ and $\mathbf{P}$. Indeed, these commute with $\mathbf{K}_\text{grav}$ according to the Poisson brackets \eqref{2.25poisson}. One also checks this explicitly using coordinates $(k,\ell)$, via the expressions \eqref{2.20hgrav} and \eqref{2.24bp}. This is the usual quantum gravity operator algebra in the absence of matter because \eqref{2.33} acts trivially on physical states in that case.

As in \cite{Chen:2024rpx}, we imagine we have some classical observer worldline in our quantum spacetimes in order to break the U$(1)$ isometry, and determine the spatial location of the operators - but we dress our local matter observables to the quantized background, and \emph{not} the observer. But even in the worst case, we can simply average the operators in the spatial direction, and most (if not all) of our story remains the same. It may be interesting to break the U$(1)$ isometry using quantum gravity, but this is beyond our scope. 

This construction is an example of construction ``relational observables'' or using ``quantum reference frames'' \cite{dewitt1967quantum,page1983evolution}.\footnote{For some more recent discussions see \cite{DeVuyst:2024uvd,Freidel:2025ous,Wei:2025guh,Akers:2025ahe,Abdalla:2025gzn,Ciambelli:2024swv}.} One way to understand this is that the gravitational constraint \eqref{2.30constraint} is nothing but a Schrodinger equation for matter states with $S$ as a physical time variable, upon writing $K_\text{grav}=\i\hbar \d/\d S$. Statement such as ``the matter state is $\psi_\text{matter}$ at time $S$'' translate into the state $\ket{\psi_\text{matter}}\otimes \ket{S}$ (with appropriate group averaging \cite{Ashtekar:1995zh,Giulini:1998rk} to render the state physical).\footnote{Some examples were worked out in section 4.3 of \cite{Blommaert:2025eps}. For more background material on group averaging, see \cite{Marolf:1996gb,Marolf:2008hg,Marolf:2000iq,Giulini:1999kc}.} To place a matter operator ``at time $S$'' one considers the operator $a\otimes \ket{S}\bra{S}$ (also with appropriate group averaging). A way of interpreting the constraint \eqref{2.30constraint} and the associated physical operators \eqref{2.34operators} is that only the combination $s+S$ is physical. For instance we have the following equivalence:
\begin{equation}
    a\otimes \ket{S}\bra{S}\sim a(S)\otimes \ket{0}\bra{0}
\end{equation}
Either we consider the slice with time $S$, and insert a ``simple'' local matter operator $a$. Or, equivalently, we consider the $S=0$ maximum volume slice of the geometry, and insert a matter operator $a(S)$ that is made by evolving into the future for a time $S$ with $K_\text{matter}$, inserting the operator $a$, and then evolving back into the past to the maximum volume slice. These two statements are physically equivalent.

\subsection{States and algebra}\label{sect:2.4ch}

To discuss the algebra of observables and the gravitational entropy, we must specify our Hilbert space. Indeed, any physical algebra must be ``completed'': operators whose matrix elements are approximated arbitrarily well by those already in the algebra should also be included. Even if we start out with the same set of well-motivated observables (such as the ones described above in (\ref{2.33}) and (\ref{2.34operators})), different choices of states can result in different algebras upon completion. An example is the infinite spin chain: at zero temperature, the entropy is well defined; at infinite temperature, only the entropy difference is well-defined; and at finite temperature, neither of them is!\footnote{See section 6 of \cite{Witten_note_ent}.} So, to discuss the algebra, we had better be clear about our states and Hilbert space of JT closed universe. 

As a spoiler, given our choice of states, we will conclude that the matter algebra on closed universe is of Type III. This might seem impossible. As explained in Section 2.4 of \cite{witten2022doesquantumfieldtheory}, QFT on closed universe usually admits a unique Hilbert space, and the algebra is of Type I.\footnote{The heuristic argument goes as follows. Because the Cauchy slices are compact, there are finitely many long-wavelength modes above some UV cutoff $\lambda>\lambda_\text{UV}$. These have a unique Hilbert space representation (up to isomorphism) due to the Stone-von Neumann theorem. For UV modes (below the cutoff), the spacetime resembles Minkowski. Because Minkowski is stationary, there is a canonical separation of creation and annihilation operators, and thus a canonical Hilbert space for the UV modes as well. As a consequence, the natural algebra of observables, is the algebra of bounded operators acting on such a Hilbert space, which is always of Type I.} However, the argument for why the algebra is type I breaks down when we have an infinite number of matter fields. This is the case for holographic CFTs at large $c$, analogous to the large $N$ limit in Liu and Leutheusser \cite{Leutheusser:2021frk}. We consider such holographic CFTs. After coupling to JT gravity, this type III algebra becomes a Type II$_{\infty}$, with meaningful gravitational entropy. 

Readers who lack interest in this derivation may skip to equation \eqref{eq254}.

We stress that our construction of gauge invariant observables makes perfect sense in the ``ordinary'' type I case. The examples with type III matter are aimed in part to prepare the reader for the discussion on subregions in section \ref{sect:4obssubr}.

Avoiding as much as possible von Neumann algebra jargon, we first explain why the thermal density matrix on the CFT Hilbert space does not exists. This means the algebra is not of type I. We will then walk through the TFD construction. Finally, we will explain why our relational observables (\ref{2.33}) and (\ref{2.34operators}) form a crossed product algebra, whose commutant is also sensible.

As explained around (\ref{2.8braket}) and (\ref{eq:betamat}), our 2d CFT thermal state is at an inverse temperature $2\pi$ with respect to conformal time, but its correlation functions are those on a torus with ``inverse temperature'' $\beta_{\rm matter}=\frac{2\pi}{B}$.\footnote{The correlators are torus blocks. In the definition of the TFD state \eqref{eq:tfdexpv} we will define the algebra via these Euclidean torus correlators. The HP transition is a feature of these torus correlators and occurs at $B=2\pi$. Since these correlators determine the algebra, we will obtain a type I algebra for small universes $B<2\pi$ and a type III algebra for large universes $B>2\pi$. This is a consequence of conformal correlators being scale invariant - such that a small universe with Euclidean time $2\pi$ is equivalent to a universe of spatial size $2\pi$ with large inverse temperature (large Euclidean time direction).} To realize this state it is natural to try the CFT Hilbert space $\mathcal{H}_{\rm matter}$ on closed universe and represent the thermal state as the thermal density matrix acting on it:
\begin{align}\label{eq:rhoth}
    \rho_{\rm th} = \frac{1}{Z(\beta)}\sum_i e^{-\beta E_i}\ket{E_i} \bra{E_i}.
\end{align} On this Hilbert space, the expectation values would have to be computed via:
\begin{align}\label{eq:expth}
    \langle a\rangle_\beta = \frac{1}{Z(\beta)}\tr_{\mathcal{H}_{\rm matter}} e^{-\beta H} a.
\end{align} 
However, as we now explain, neither of these two equations makes sense for large enough closed universes (and a 2d holographic CFT), because of the explicit appearance of $Z(\beta)$, which diverges. 

As a preparation, recall the familiar case of $\mathcal{N}=4$ SU$(N)$ SYM on the cylinder $\d s^2 = -\d t^2+\d x_i^2$. As explained for instance in \cite{witten2022doesquantumfieldtheory}, the partition function $Z(\beta)$ is expected to diverge (in the appropriate large $N$ limit for holography) when the temperature is above the Hawking-Page temperature, $\beta <\beta_{\rm HP}$. So, neither the thermal density matrix nor the RHS of (\ref{eq:expth}) makes sense at high enough temperature. To construct a sensible Hilbert space that realizes the thermal state, we need to avoid the appearance of $Z(\beta)$ throughout the construction. This is achieved by going to its canonical purification, known as the
TFD $\ket{\psi_{\rm TFD}}_\beta$. If the partition function were finite $Z(\beta)<\infty$ the TFD would simply be
\begin{align}\label{eq:psitfd}
    \ket{\psi_{\rm TFD}}_\beta = \frac{1}{\sqrt{Z(\beta)}}\sum_i e^{-\beta E_i/2}\ket{E_i} \otimes \ket{\overline{E_i}}\,.
\end{align} 
This lives in the double copy of the original Hilbert space, $\mathcal{H}_{\rm matter}\otimes\overline{\mathcal{H}}_{\rm matter}$. Of course, if the partition function were finite, (\ref{eq:rhoth}) would have been well-defined, and we wouldn't have needed the TFD. 

The TFD becomes \emph{useful} when $Z(\beta)=\infty$ (so that equation (\ref{eq:psitfd}) is meaningless). In this case, the TFD state remains well-defined as map from operators to $O(1)$ number, namely the thermal expectation values of these operators:
\begin{align}\label{eq:tfdexpv}
\langle \psi_{\mathrm{TFD}}|\,\cdot\,|\psi_{\mathrm{TFD}}\rangle
  :\quad 
a \in  \mathcal{A}_{\mathrm{QFT}} &\longmapsto \langle a\rangle_\beta\in \mathbb{C} .
\end{align}
One then builds a vector space of states by acting with all combinations of operators on the TFD. This is a vector space, as the algebra of operators is a vector space by definition. The inner product between states $a\ket{\psi_{\rm TFD}}$ and $b\ket{\psi_{\rm TFD}}$ is defined to be $\langle a^\dagger b\rangle_\beta$. This space is completed to a Hilbert space $\mathcal{H}_{\rm TFD}$. The key point is that none of the above steps involves explicit factor of $Z(\beta)$, so $\mathcal{H}_{\rm TFD}$ is well-defined \emph{even} when the partition function diverges! More details can be found in for instance \cite{Witten_note_ent,witten2022doesquantumfieldtheory}.

\begin{figure}
    \centering
    \begin{tikzpicture}[baseline={([yshift=-.5ex]current bounding box.center)}, scale=0.7]
 \pgftext{\includegraphics[scale=1]{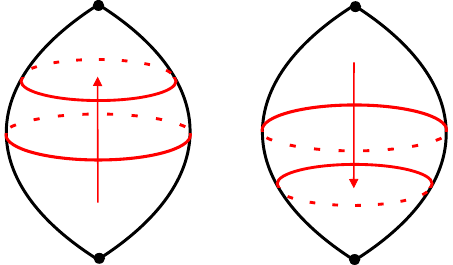}} at (0,0);
    \draw (-2.2,3.6) node {closed universe};
    \draw (2.2,3.6) node {reservoir};
    \draw (-2.2,-2.6) node {\color{red}$s=-\infty$};
    \draw (-2.2,2.6) node {\color{red}$s=+\infty$};
  \end{tikzpicture}
    \caption{The CFT thermal state prepared by the bra-ket wormhole can be realized as a thermofield double on two copies of a closed universe. We call the original universe ``closed universe'' and the auxiliary copy the ``reservoir''. We only turn on gravity for the first copy. The red arrow indicates the flow of modular time.}
    \label{fig:tfd}
\end{figure}
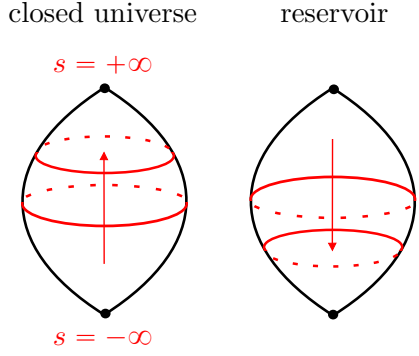

How does this apply to the algebra of observables in AdS$_2$ closed universes? The analogue of single trace operators of $\mathcal{N}=4$ SYM are light primaries of a (bulk) holographic 2d CFT. The analogue of the TFD is the state prepared by the bra-ket wormhole contour, whose correlation functions were given in \eqref{2.17G}. More precisely, \eqref{eq:tfdexpv} in the TFD construction is replaced by the (thermal) conformal correlator with torus modular parameter $\beta_\text{matter}=2\pi/B$:
\begin{equation}\label{eq:braketdef}
\bra{\psi_\text{bra-ket}}\widetilde{\mathcal{O}}_\Delta (s_1)\widetilde{\mathcal{O}}_\Delta(s_2)\ket{\psi_\text{bra-ket}}=G_{\text{torus}\,\Delta}(s_1,s_2)\,.
\end{equation} 
This satisfies KMS as discussed in equation \eqref{2.15kms}, for $s\to s+2\pi \i$. From now on we call $\ket{\psi_\text{bra-ket}}\to \ket{\psi_\text{TFD}}$. The point is that, for $B>2\pi$, the holographic CFT is above the HP temperature, so that the torus partition function diverges $Z_\text{torus}(2\pi/B)=\infty$. As we pointed out this means that $\ket{\psi_\text{bra-ket}}$ is not in the closed universe Hilbert space and instead should be viewed in the formal spirit of the TFD, thus  realized as an entangled state with another closed universe (where time runs backwards). See figure \ref{fig:tfd}. In the language of von Neumann algebras, the non-existence of thermal density matrices implies that the algebra is not of Type I. In fact, following Liu-Leutheusser \cite{Leutheusser:2021frk}, our algebra of large closed universe $\mathcal{A_{\rm matter}}$ is of Type III!\footnote{This is because the light primaries of a 2d holographic CFT above the Hawking-Page transition are dual to bulk matter fields on an exterior wedge of 3d BTZ black hole (by holography). Because the wedge is a subregion in QFT, its associated algebra $\mathcal{A}_{\rm matter}$ is of Type III$_1$.} We like type III matter for entropic reasons (bear with us). 

We now turn on gravity, but only for one of the closed universe. The motivation for doing so is that the other closed universe is simply auxiliary. It is a mathematical byproduct of the TFD construction. \footnote{In our case of JT gravity, we do not need a second copy of gravitational modes to realize the states. It should be emphasized, though, that this could be an artifact of 2D dilaton gravity, and that an entangled pair of gravitational closed universes is interesting on its own right. We leave that for future study.} From hereon we refer to the gravitational universe as the ``closed universe'' and the non-gravitational one as ``reservoir.''  Prior to imposing the gravitational constraint (\ref{2.30constraint}), the Hilbert space reads:
\begin{align}\label{eq:hglobal}
    \mathcal{H} = \mathcal{H}_{\rm JT}\otimes\mathcal{H}_{\rm TFD}\,,
\end{align} where the former is the pure JT Hilbert space. As explained around (\ref{2.25poisson}), this (unconstrained) Hilbert space consist of wavefunctions of, say, the maximal size of universe $B$ and the conformal time ``energy'' $K$:
\begin{align}
    \ket{\psi}_{\rm JT} = \int_{0}^{+\infty}\d B\,\psi(B)\ket{B}\otimes \int_{-\infty}^{+\infty}\d K\, \phi(K)\ket{K}\quad\in\quad \mathcal{H}_{\rm JT}=L^2(\R)\otimes L^2(\R)\,.
\end{align} 
Here, $\psi(K)$ and $\phi(B)$ are normalizable distributions, for instance semiclassical Gaussians. When matter is absent, the constraint (\ref{2.30constraint}) simply kills the $K$ mode, so invariant wavefunctions are $\psi(B)$, agreeing with \cite{Held:2024rmg}. With matter, the full state becomes:
\begin{align}
\ket{\psi}=\int_{0}^{+\infty}\d B\,\psi(B)\ket{B}\otimes \int_{-\infty}^{+\infty}\d K\, \phi(K)\ket{K}\otimes \ket{\psi_\text{matter}},\quad\ket{\psi_{\rm matter}}\in\mathcal{H}_{\rm TFD}\,,\label{2.38state}
\end{align} where, as a reminder, the matter state $\ket{\psi_{\rm matter}}$ can be approximated arbitrarily well by finite excitations acting on the TFD $\ket{\psi_{\rm TFD}}$. The constraint will be imposed at the level of algebra later. This concludes our discussion of the Hilbert space. 

Now we move on to the algebra. Before imposing the constraint (\ref{2.30constraint}), observable algebra on closed universe is
\begin{align}
    \mathcal{A}_0 = \mathcal{B}(L^2(\R)\otimes L^2(\R))\otimes\mathcal{A}_{\rm matter}.
\end{align} 
The physical subalgebra $\mathcal{A}$ is the completion (the double commutant) of the observables of section \eqref{sect:2.3closedobs}:
\begin{align}\label{eq:ainvt}
    \mathcal{A} = \text{completion of }\{\text{bounded functions of } \mathbf{B}\text{ and } \mathbf{P}\}\otimes\{\text{bounded functions of } a(s+\mathbf{S}) \text{ and } \mathbf{K_{\rm grav}})\}\,.
\end{align} 
The double commutant (the completion) is taken with respect to \eqref{2.38state}. Because the constraint \eqref{2.30constraint} sets $K_{\rm grav}$ equal to (minus) the matter modular Hamiltonian $-K_{\rm matter},$ the second part of the gauge-invariant algebra \eqref{eq:ainvt} takes exactly the form of crossed product algebra of $\mathcal{A}_{\rm matter}$! It is now tempting to argue that the unconstrained algebra $\mathcal{A}_0$ is of Type III, and the gauge-invariant algebra $\mathcal{A}$ is of Type II, which would mean that entropy (difference) is well-defined. There is, however, one technical subtlety. 

Recall that in arguing that $\mathcal{A}_{\rm matter}$ is of Type III, we needed the size of closed universe to be large enough: $B>2\pi.$ But once gravity is turned on, we are forced to consider states whose $B$ is \emph{below} this threshold, or, more generically, states whose $B$ can fluctuate above \emph{and} below the threshold, i.e., the wave function $\psi(B)$ have support on both sides of $B=2\pi.$ We emphasize that this makes neither the algebra (\ref{eq:ainvt}) nor the Hilbert space (\ref{eq:hglobal}) any less well-defined. This merely complicates the discussion on the type of von Neumann algebra\footnote{At this point, a reader (especially those who haven't worked on semiclassical algebras) might wonder why we are obsessed with the type of algebra, and that's a good point. Perhaps we shouldn't! Knowing the type of algebra is often convenient because there are some off-the-shell theorems at our disposal which were invented by mathematicians half-a-century ago, and which have led to interesting physics insights e.g. \cite{Witten:2021unn,Chandrasekaran:2022eqq,Chandrasekaran:2022cip}. But perhaps it is time to bravely study the algebra that shows up naturally in quantum gravity, \eqref{eq:ainvt} is the algebra that shows up in JT gravity.}. 

To connect with the usual Type II story, we restrict ourselves to a fixed $B>2\pi$ sector. The Hilbert space becomes:
\begin{align}
    L^2(\R)\otimes \mathcal{A}_{\rm matter}\,,
\end{align} where the first factor is the space of $K$ wavefunctions. The gauge-invariant algebra is
\begin{align}
    \mathcal{A}=\text{completion of }\{\text{bounded functions of } a(s+\mathbf{S}) \text{ and } \mathbf{K_{\rm grav}})\}\,.\label{eq249}
\end{align} 
For $B>2\pi$, the matter algebra is of Type III, and the gauge-invariant algebra $\mathcal{A}$ is of Type II$_\infty$! The entropy difference is well-defined and will be discussed in section \ref{sect:2.4closedentropy}.

As a sanity check, we show that the commutant of (\ref{eq:ainvt}) is indeed the algebra of observables acting on the reservoir (the universe on the right in figure \ref{fig:tfd}). This is straightforward because the commutant of the crossed product algebra is well-known. Denote the matter modular Hamiltonian for closed universe $+$ reservoir by $K_{\rm mod}$. This hence decomposes as:
\begin{align}\label{eq:Kmod}
    K_{\rm mod} = K_{\text{mat}\,\text{CU}}-K_{\text{mat}\,\text{R}}\,.
\end{align} 
Here CU denotes ``closed universe'' and R denotes ``reservoir''. With this notation, the constraint (\ref{2.30constraint}) reads:
\begin{align}\label{eq:newconstr}
    K_{\rm grav}+K_{\text{mat}\,\text{CU}}=0\,.
\end{align} 
It is well-known that if the crossed product algebra is generated by $e^{\i S K_{\rm grav}} a e^{-\i S K_{\rm grav}}$ and $K_{\rm grav}$, namely \eqref{eq249} then the commutant is generated by $a'$ and $K_{\rm mod}+K_{\rm grav},$ where $a'$ are operators that commute with all $a$'s. This is explained for instance around (3.6) of \cite{Witten:2021unn} for the case of two-sided black hole; our algebra of closed universe and the commutant is like their left algebra $\mathcal{A}_L$ and the right one $\mathcal{A}_R$ in their equation (3.6). Therefore:
\begin{align}
    \mathcal{A}' = \text{double commutant of }\{\text{bounded functions of }a'(s) \text{ and }K_{\rm mod}+K_{\rm grav}\}\,.\label{2.52al}
\end{align} 
Plugging in the expression for the modular Hamiltonian \eqref{eq:Kmod} and the constraint \eqref{eq:newconstr}, one sees that the second operator appearing in \eqref{2.52al} is (minus) the conformal time Hamiltonian of the reservoir! 
\begin{align}
    K_{\rm mod}+K_{\rm grav} = -K_{\text{mat}\,\text{R}}\,.
\end{align}
Obviously $a'(s)$ are reservoir matter operators, so the commutant algebra is indeed exactly the matter algebra on the reservoir (which is already gauge invariant because the the reservoir is non-gravitational) plus its Hamiltonian - as expected.

For concreteness, let us mow look at an example of physical observables in our framework. Consider the the dressed two-point correlation function in the state,
\begin{align}
    \ket{\psi}=\int_{0}^{+\infty}\d B\,\psi(B)\ket{B}\otimes \int_{-\infty}^{+\infty}\d K\, \phi(K)\ket{K}\otimes \ket{\psi_\text{TFD}}.\label{eq254}
\end{align} Namely, the matter state is simply the TFD. As a reminder, the TFD two-point correlator equals:
\begin{equation}
\bra{\psi_\text{TFD}}\widetilde{\mathcal{O}}_\Delta (s_1)\widetilde{\mathcal{O}}_\Delta(s_2)\ket{\psi_\text{TFD}}=G_{\text{torus}\,\Delta}(s_1,s_2\rvert \tau(B))
\end{equation}
Here we explicitly indicated the $B$ dependence of matter correlators, which enters through the modulus $\tau(B)=\i/B$ of the Euclidean torus used to prepare the matter state. We then compute a gauge-invariant matter two-point correlator in this state as follows:
\begin{align}
\bra{\psi}\widetilde{\mathcal{O}}_\Delta(s_1+\mathbf{S})\widetilde{\mathcal{O}}_\Delta(s_2+\mathbf{S})\ket{\psi}&=\int_0^\infty \d B\,\abs{\psi(B)}^2\int_{-\infty}^{+\infty}\d K_1\,\phi(K_1)^*\int_{-\infty}^{+\infty}\d K_2\,\phi(K_2)\int_{-\infty}^{+\infty}\d S\nonumber\\&\qquad\qquad\qquad e^{\i S(K_1-K_2)/\hbar}\, \bra{\psi_\text{TFD}}\widetilde{\mathcal{O}}_\Delta(s_1+S)\widetilde{\mathcal{O}}_\Delta(s_2+S)\ket{\psi_\text{TFD}}\nonumber\\&=\int_0^\infty \d B\,\abs{\psi(B)}^2\,G_{\text{torus}\,\Delta}(s_1,s_2\rvert \tau(B))\,.\label{2.542point}
\end{align}
In the second equality we used translation invariance of the matter correlator, as discussed below \eqref{2.12torus}, showing that the matter matrix element on the second line is independent of $S$. Then the $S$ integration produces $\delta(K_1-K_2)$, and we normalized the integral of $\abs{\phi(K)}^2$ to unity. The final result is the ordinary matter correlator integrated over $B$. This eliminates light-cone singularities due to winding around the compact spatial direction, which arise at $B$-dependent times $s_1-s_2\in B \mathbb{Z}$.

One might have expected that dressing to a dynamical geometric time variable results in additional smearing of the correlator in Lorentzian times, as pointed out by Witten around equation (3.40) in \cite{Witten:2023xze}. In this example, that does \emph{not} happen. One can imagine other correlators which \emph{do} result in Lorentzian time smearing, though. For instance, a similar calculation for a natural gauge-invariant observable gives:
\begin{equation}
\bra{\psi}\mathcal{O}_\Delta(s_1+\mathbf{S})\,\abs{\phi(\mathbf{K}_\text{grav})}^2\,\mathcal{O}_\Delta(s_2+\mathbf{S})\ket{\psi}=\int_0^\infty \d B\,\abs{\psi(B)}^2\int_{-\infty}^{+\infty}\d S\,\abs{\phi(S)}^2\,G_{\text{torus}\,\Delta}(s_1-s_2+S\rvert \tau(B))\,.\label{2.55smeared}
\end{equation}
Here
\begin{equation}
    \phi(S)=\int_{-\infty}^{+\infty}\d K\,\abs{\phi(K)}^2\,e^{\i K S/\hbar }\,.
\end{equation}
One could imagine the two factors $\abs{\phi(\mathbf{K}_\text{grav})}^2$ (one in between the matter operators, one in the definition of the state itself) to arise from inserting a completeness relation for a non-ideal clock $\mathbf{K}_\text{grav}$. One could speculate that including topology changing effects in the gravitational path integral, for instance, may lead to a Hilbert space where $\mathbf{K}_\text{grav}$ effectively has some maximal value - much like the size of a Cauchy size between two asymptotic AdS boundaries \cite{Iliesiu:2021ari,Iliesiu:2024cnh}. This might result in a sharply peaked $\phi(S)$, with width $\sim 1/K_\text{max}$. It would be interesting to understand if the gravitational path integral provides with a mechanism such as this to time-average matter correlators. This might be some mechanism to resolve the (Milne-type) Big-Bang singularity. One might worry if such correlators on the Milne geometry are even finite after time smearing, because of the repeated lightcone singularities from winding around the spatial circle. As we show in appendix \ref{app:finitecorr}, they are indeed finite.

Before proceeding, we make one technical remark. Strictly speaking in the infinite $c$ limit, except for $K_{\rm mod}$, none of the operators in \eqref{eq:Kmod} and \eqref{eq:newconstr} are well-defined (just like one-sided boosts for a black hole). The fluctuations in $K_{\rm grav}$ diverges, for instance. The same issue shows up for black holes \cite{Chandrasekaran:2022eqq}. The solution there is to go to the microcanonical ensemble, so that the fluctuation in clock Hamiltonian (the right ADM mass in the black hole case) is by definition finite. One could imagine that we can do something similar by considering a micro-canonical version of the bra-ket wormhole state. We leave a more detailed investigation of these subtleties for future study.

\subsection{Type II gravitational entropy (only with doubly holographic conformal matter)}\label{sect:2.4closedentropy}
We constructed gauge-invariant local conformal matter operators by dressing to the extrinsic curvature $K$ of Cauchy slices, and showed that they form a Type II$_\infty$ algebra \eqref{eq249}. One reason for the increased recent interest in understanding such gauge-invariant observables is that, in certain cases, constructing such observables has shed light on gravitational entropy. In particular, it has resulted in an improved understanding of the area term in the generalized entropy attributed to horizons - like the cosmological horizon of the dS static patch and black hole horizons in AdS \cite{Chandrasekaran:2022cip,Chandrasekaran:2022eqq,Kudler-Flam:2023qfl,Jensen:2023yxy}. In these cases, the algebra of local matter operators (prior to coupling to gravity) is a type III von Neumann algebra. Such type III algebras have \emph{no} well defined notion of entropy. The coupling to gravity, leading to a gauge-invariant operator algebra that is morally build out of operators of the types \eqref{2.33} and \eqref{2.34operators}, results in a type II von Neumann algebra \cite{Witten:2021unn} - which \emph{does} have a notion of entropy associated with states.

 Although an area term in the entropy is expected, this cannot be true for closed universes, simply because there are no boundaries. Therefore, the gravitational entropy (if nontrivial) is necessarily a bulk gravitational quantity. This discussion is aimed in part at preparing readers for a similar conclusion in our discussion of gravitational subregion entropy in section \ref{sect:4.3entropy} - where entropy isn't a boundary term either. The present discussion should clarify that this is uncontroversial, and in some cases \emph{unavoidable}. In the discussion section \ref{sect:concl}, we speculate whether this gravitational entropy might retain some meaning in cases where the matter entropy is type I. There are indeed cases where one might expect a non-zero gravitational entropy for closed universes, even for type I matter. This issue was raised in section 6 of \cite{Blommaert:2025bgd}. Unfortunately, we will not resolve this puzzle.

The computation of gravitational entropy is quite technical. We will assume the reader is familiar with the formulas in e.g. section 3 of CLPW \cite{Chandrasekaran:2022cip}. The first step is to construct the ``trace'' on operators in the II$_\infty$ algebra. Intuitively, the ``tracial state'' is
\begin{equation}
    \rho_\text{NB}=e^{-2\pi \mathbf{K}_\text{grav}/\hbar}\rho_\text{th}=e^{-2\pi (\mathbf{K}_\text{grav}+K_\text{matter})/\hbar}\label{2.43trace}\,,
\end{equation} where $\rho_{\rm th}$ is, as in previous sections, the thermal state prepared by CFT path integral along the bra-ket wormhole contour. 
Indeed, using matter KMS \eqref{2.15kms} and the definition of $a(s+\mathbf{S})$ in \eqref{2.34operators}, one finds that physical operators indeed commute with this tracial state:
\begin{equation}
    e^{-2\pi \mathbf{K}_\text{grav}/\hbar}\rho_\text{th}\, a(s+\mathbf{S})= a(s+\mathbf{S})\,e^{-2\pi \mathbf{K}_\text{grav}/\hbar}\rho_\text{th}\,.\label{2.44com}
\end{equation}
This results from the fact that $\rho_\text{NB}$ is a function of the constraint $\mathbf{K}_\text{grav}+K_\text{matter}$, with which physical operators by construction commute. The state \eqref{2.43trace} is uniquely determined, given the choice of matter state, by this requirement, and is therefore the natural gravitational dressing of $\rho_\text{th}$.

Although \eqref{2.43trace} is intuitively correct, the more refined definition of the tracial state $\psi$ is actually \cite{Chandrasekaran:2022cip}:
\begin{equation}
    \tr(a)=\int_{-\infty}^{+\infty}\d K_1\,e^{-\pi K_1/\hbar}\bra{K_1}\otimes \bra{\psi_\text{TFD}}a \ket{\psi_\text{TFD}}\otimes \int_{-\infty}^{+\infty} \d K_2\,e^{-\pi K_2/\hbar}\ket{K_2}=\bra{\psi_\text{trace}}a \ket{\psi_\text{trace}}\,,\label{2.44trace}
\end{equation}
where
\begin{equation}
    \ket{\psi_\text{trace}}=e^{-\pi \mathbf{K}_\text{grav}/\hbar}\ket{0}\otimes \ket{\psi_\text{TFD}}\,.
\end{equation}
The reason for this minor tweaking of \eqref{2.43trace} is that the latter suffers from divergences since for instance $\bra{K_1}f(\mathbf{K}_\text{grav})\ket{K_1}=\delta(K_1-K_2)f(K_1)$ would produce an overall $\delta(0)$. The proof that \eqref{2.44trace} is a trace if obvious for functions of $\mathbf{K}_\text{grav}$. One furthermore shows\footnote{Introducing several completion relations one expands
\begin{align}
    \tr(a(\mathbf{S})b(\mathbf{S}))&=\int_{\infty}^{+\infty} \d K_1 \d K_2\d K\int_{-\infty}^{+\infty} \d S_1 \d S_2\,e^{-\pi(K_1+K_2)/\hbar+\i S_1(K_1-K)/\hbar+\i S_2(K-K_2)/\hbar}\bra{\psi_\text{matter}}a(S_1)b(S_2)\ket{\psi_\text{matter}}
\end{align}
Then one uses matter KMS $\bra{\psi_\text{matter}}a(S_1)b(S_2)\ket{\psi_\text{matter}}=\bra{\psi_\text{matter}}b(S_2)a(S_2+2\pi\i)\ket{\psi_\text{matter}}$. One then does the change of integration values (and integration contours, which is allowed because KMS also assumes analyticity of matter correlators) $S_2=P_1, S_1+2\pi \i=P_1$ and furthermore $Q_1=-K+K_1-K_2, Q_2=-Q, Q=K_1-K_2$, which brings the above expression into the following form:
\begin{align}
    \tr(a(\mathbf{S})b(\mathbf{S}))&=\int_{\infty}^{+\infty} \d Q_1 \d Q_2\d Q\int_{-\infty}^{+\infty} \d P_1 \d P_2\,e^{-\pi(Q_1+Q_2)/\hbar+\i P_1(Q_1-Q)/\hbar+\i P_2(Q-Q_2)/\hbar}\bra{\psi_\text{matter}}b(P_1)a(P_2)\ket{\psi_\text{matter}}\nonumber\\ &=\tr(b(\mathbf{S})a(\mathbf{S}))\,.
\end{align}
We thank Jonah Kudler-Flam, Erez Urbach and Edward Witten for discussions/collaboration on this point.
}
\begin{equation}
    \tr(a(s_1+\mathbf{S})b(s_2+\mathbf{S}))=\tr(b(s_2+\mathbf{S})a(s_1+\mathbf{S}))\,.
\end{equation}
Notice that these simple examples do not result in finite correlation functions, because of a divergence in the $K$ integral. This is common for type II$_\infty$ algebras. Finite correlators can be obtained for instance by considering, following equation \eqref{2.38state}, expectation values of $a(s+\mathbf{S})$ in states with some normalizable gravity wavefunctions $\psi(K)$ (in section \ref{sect:2.4ch} we instead used the notation $\phi(K)$):
\begin{equation}
    \ket{\psi}=\psi(\mathbf{K}_\text{grav})\ket{0}\otimes \ket{\psi_\text{TFD}}\,.\label{2.50state}
\end{equation}
The trace \eqref{2.44trace} associates with these states the density matrices $\rho_{\text{alg}\,\psi}$ defined through
\begin{equation}
    \bra{\psi}a\ket{\psi}=\tr(\rho_{\text{alg}\,\psi} a)\quad\to \quad \rho_{\text{alg}\,\psi}=e^{2\pi \mathbf{K}_\text{grav}/\hbar}\,\abs{\psi(\mathbf{K}_\text{grav})}^2\,.\label{2.51rho}
\end{equation}
We introduced the subscript alg, to distinguish from the more naive notion of density matrices used in \eqref{2.43trace}.

To understand gravitational entropy, we consider semiclassical gravitational states, so for instance coherent state wavefunctions of $\mathbf{K}_\text{grav}$. For such states we can approximately say what the geometry is, and therefore one might except a geometric answer to come out of a quantum mechanical calculation.\footnote{A general characterizations of semiclassical states were discussed for instance in \cite{DeVuyst:2024uvd} (section 6) and \cite{Chandrasekaran:2022cip} (section 3), but coherent states are intuitive examples.} We consider states with wavefunctions very sharply peaked around some value $K$, which we denote by:
\begin{equation}
    \ket{\psi_K}=\psi(\mathbf{K}_\text{grav}-K)\ket{0}\otimes \ket{\psi_\text{TFD}}\,.
\end{equation}
Assuming $\ket{\psi_\text{matter}}$ is normalized we may compute the vN entropy associated with this state as follows
\begin{align}
    S(\rho_{\text{alg}\,K_1})=-\tr(\rho_{\text{alg}\,K_1}\log \rho_{\text{alg}\,K_1})&=-\int_{-\infty}^{+\infty}\d K\,\abs{\psi(K-K_1)}^2(2\pi K/\hbar+\log \abs{\psi(K-K_1)}^2)\nonumber\\
    &= -2\pi K_1/\hbar-\int_{-\infty}^{+\infty} \d K \abs{\psi(K)}^2\log \abs{\psi(K)}^2=-2\pi K_1/\hbar+S_\psi
\end{align}
In the second line we used that the distribution $\abs{\psi(K-K_1)}^2)$ is unit normalized and is sharply peaked around $K_1$. The geometric component is isolated by investigating entropy differences between two such semiclassical states with identical $\psi$ (but different center of the distribution):
\begin{equation}
    \boxed{\delta S=-\frac{2\pi}{\hbar}\delta K_\text{grav}\,}\label{2.54entropyv1}
\end{equation}
Only such entropy differences are meaningful using type II traces anyway. This equation is the analog of equation (3.18) in CLPW \cite{Chandrasekaran:2022cip}, which contains the results of a similar calculation which also compares different matter states (for simplicity we considered identical matter states). That entropy (differences) is minus gravitational energy (differences) is a universal feature in these types of gravitational crossed product constructions (for another intuitive example consider equation (4.6) in \cite{Witten:2023xze}).

We stress that this contribution arises because of naively counterintuitive facts about type II traces and density matrices. Firstly, the trace \eqref{2.44trace} contains a factor $e^{-2\pi \mathbf{K}_\text{grav}/\hbar}$, which is funny as compared to our usual notion of tracing over some Hilbert space. Secondly, the definition of the density matrix \eqref{2.51rho} associated with the state \eqref{2.50state} contains a factor $e^{+2\pi \mathbf{K}_\text{grav}/\hbar}$ which is unexpected from our usual notion of density matrices. Type II algebras are funny.\footnote{Schematically one relates this to the more naive trace $\Tr$ (which does not have this extra exponential in the $K$ integral) using
\begin{equation}
    -\tr(\rho_{\text{alg}\,K}\log \rho_{\text{alg}\,K})=-\Tr(\rho_K \log \rho_\text{NB}^{-1}\rho_K)=-S(\rho_K\rvert\rvert\rho_\text{NB})\,,
\end{equation}
where $\rho_K=\rho_\text{NB}\rho_{\text{alg}\,K}$ and using \eqref{2.43trace}. This is the statement that type II von Neumann entropy equals minus the relative entropy comparing with the tracial (no-boundary) state \cite{Witten:2023xze,Blommaert:2025bgd}.
}

We would like to find the geometric meaning of \eqref{2.54entropyv1}. What replaces $\delta S=\delta A/4 G$ for the (classical) dS static patch \cite{Chandrasekaran:2022cip}? We claim that the correct answer is: the integral of the dilaton $\Phi$ over the geodesic (or maximum volume) Cauchy slice. In equations:\footnote{Recall that in the JT gravity action \eqref{2.1jtac} we used conventions $8\pi G=1$ and left $\hbar$ implicit. In more common conventions where $\hbar=1$ and $G$ remains one finds
\begin{equation}
    \delta S=\frac{1}{4 G}\oint_\text{geo}\d x\,\delta\Phi(x)\,.\label{2.70G}
\end{equation}
}
\begin{equation}
    \boxed{\delta S=\frac{2\pi}{\hbar}\oint_\text{geo}\d x\,\delta \Phi(x)\,}\label{2.57entropyformula}
\end{equation}
This is derived using the Einstein equations. Recall the JT gravity Einstein equation \eqref{2.27Eeq}, projected onto the CKV $\xi=\d/\d s$:
\begin{equation}
    \bigg\{\frac{\d^2}{\d x^2}+\tanh(s)\frac{\d}{\d s}-\frac{1}{\cosh(s)^2} \bigg\}\Phi=-\sqrt{h}n^\mu \xi^\nu T_{\mu \nu}\label{2.58tgrav}
\end{equation}
Integrating over $x$ and using the fact that the dilaton is periodic in the spatial direction $\Phi(x+B)=\Phi(B)$ one obtains\label{eq:kgravcu}
\begin{equation}
    K_\text{grav}=\oint \d x  \bigg\{\tanh(s)\frac{\d}{\d s}-\frac{1}{\cosh(s)^2} \bigg\}\Phi(x)
\end{equation}
This expression simplifies considerably on the maximal volume slice $s=0$, where $\d \Phi/\d s$ remains finite:
\begin{equation}
    K_\text{grav}=-\oint_\text{geo}\d x\,\Phi(x)\,.
\end{equation}
Combined with \eqref{2.54entropyv1} this indeed proves the gravitational entropy formula \eqref{2.57entropyformula}. For orientation, we consider aa specific example spatially homogeneous matter profiles
\begin{equation}
    \sqrt{h}n^\mu \xi^\nu T_{\mu \nu}=\frac{K_\text{matter}}{B}=-\frac{K_\text{grav}}{B}\,.
\end{equation}
The unique solution of the JT gravity Einstein equations \eqref{2.27Eeq} is then a spatially homogeneous dilaton profile:\footnote{Notice that the gravitational phase space structure is not affected by the matter configuration, as in \cite{Held:2024rmg}. The Einstein equations are linear in $\Phi$ therefore the source term (particular solution) simply adds to the general solution (quantized in phase space).}
\begin{equation}
    \Phi=\Phi_h\tanh(s)+\Phi_\text{sourced}\,,\quad \Phi_\text{sourced}=\frac{K_\text{matter}}{B} (s\tanh(s)-1)\,.\label{2.62dilsol}
\end{equation}
We did not find any relation between the gravitational entropy formula \eqref{2.57entropyformula} and changes in the dilaton profile at an extremal surface (for $K_\text{matter}>0$ there is a surface where $\d \Phi/\d s=0$). In JT gravity, the dilaton plays the role of an area, and therefore we conclude that entropy changes are not extremal area changes in this setup. In fact, the entropy \eqref{2.57entropyformula} is a bulk quantity. As we are considering homogeneous closed universes, this is perhaps not too surprising. Nevertheless, it would be interesting to gain more intuition for this expression. In section \ref{sect:4.3entropy}, we recall that a similar calculation \emph{does} result in the usual area law for the gravitational entropy of JT black holes \cite{Kolchmeyer:2023gwa,Penington:2023dql}. See also appendix \ref{sec:appxb}.

Before proceeding let us make two more comments.
\begin{enumerate}
    \item One is inclined to associate some Euclidean gravitational action with the entropy formula \eqref{2.57entropyformula}. Following \cite{Witten:2023xze,Blommaert:2025bgd} one might want to relate the trace (and expectation values of physical operators in the trace \eqref{2.44trace}) with a no-boundary \cite{hartle1983wave} path integral. By analogy with the geometry preparing the matter state \eqref{2.8braket} one would be inclined to consider the GPI for the same bra-ket wormholes. The algebraic considerations suggest that for configurations with fixed $K_\text{grav}$, we expect roughly the following Euclidean action:
    \begin{equation}
        e^{-I}=e^{-2\pi K_\text{grav}/\hbar}\,,\label{2.63ac}
    \end{equation}
    One could imagine that this comes from the phase space term in the gravitational action, associated with the symplectic from \eqref{2.25poisson}:
    \begin{equation}
        \exp(\frac{\i}{\hbar}\int_0^1 \d \tau\,K_\text{grav}\frac{\d S}{\d \tau})\,.
    \end{equation}
    Imposing the monodromy condition consociated with the bra-ket wormhole contour \eqref{2.8braket} $S(1)-S(0)-2\pi\i$ results in \eqref{2.63ac}. In quantum mechanics, indeed, entropy usually arises from evaluating the phase space terms in the on-shell action.\footnote{See for instance equation (2.20) in \cite{Blommaert:2024ydx} for another example.} It would be interesting to make this more precise.

    Such an explanation would \emph{still} not obviously be consistent with the no-boundary path integral! It was argued in \cite{Blommaert:2025bgd} that the no-boundary path integral prepares (roughly speaking) the following density matrix, which is a projector on the physical Hilbert space:
    \begin{equation}
        \delta(\mathbf{K}_\text{grav}+K_\text{matter})\,.
    \end{equation}
    However, the trace \eqref{2.44trace} does not (naively) implement such a projector. Rather, $e^{-2\pi \mathbf{K}_\text{grav}/\hbar}\rho_\text{th}$ is the unique combination which involves $\rho_\text{th}$ \emph{and} is gauge-equivalent to the identity operator, as discussed below \eqref{2.44com}. How are these two pictures consistent? Where is the projector on physical intermediate states in \eqref{2.44trace}? We speculate on a potential resolution in the concluding section \ref{sect:concl}.

    \item Amusingly, the sourced dilaton solution \eqref{2.62dilsol} blows up near the Big-Bang and the Big-Crunch $s\to\pm\infty$. The current JT gravity description likely breaks down when $\Phi_\text{break}=-S_0/4\pi$, because then wormhole corrections become unsuppressed. This happens when $s=-S_0 B/4\pi K_\text{matter}$. Thus wormholes become important close to singularities. In similar spirit to the discussion at the end of section \ref{sect:2.4ch}, one might speculate that non-perturbative effects effectively lead to a gravitational Hilbert space where $\abs{s}<s_\text{max}$, thus shielding gravitational singularities (at least in JT gravity). We leave investigating such  potential mechanisms of singularity resolution to future work.
\end{enumerate}

This concludes are discussion of gauge-invariant local observables in AdS JT gravity closed universes.

\section{Non-perturbative subregions}\label{sect:3diamons}
In this section, we study the quantum mechanics of causal diamonds, in pure JT gravity. In section \ref{sect:4obssubr}, we include conformal matter, and repeat our construction of gauge-invariant local operators of section \ref{sect:2obcl}. We built up this work such that the analysis and results of section \ref{sect:4obssubr} seem like obvious consequences of combining the material of section \ref{sect:2obcl} and of the current section.

In \textbf{section \ref{sect:3.1classicaldiamonds}} we define our subregions of interest. We will be studying subregions of the two-sided black hole geometry. We define our subregions by specifying the value of the dilaton $\Phi$ at two endpoints. We then consider the causal diamond associated with such a spacelike subregion. For fixed $\Phi_\text{bdy}$ there is a two-dimensional phase space of causal diamonds, labeled by the geodesic size $B$ of each subregion, and its conjugate $P$. As in section \ref{sect:2.2Ktime}, there is a third coordinate $S$ on the space of subregions, associated with the extrinsic curvature (or the conformal time) of a chosen slicing of the diamond. This coordinate $S$ represents a diffeomorphism of the subregion, and is redundant \cite{Jensen:2023yxy}. Its conjugate variable ($K_\text{grav}$) is the gravitational constraint generating CKV flow in these diamonds. In \textbf{section \ref{sect:3.2quantumdiamonds}} we discuss this constraint, various coordinates on the physical phase space, and their relation with quantization of JT gravity at ``finite cutoff'' \cite{Iliesiu:2020zld,Griguolo:2025kpi}. Then, in \textbf{section \ref{sect:3.3assdiamond}}, we relate with ``ordinary'' JT gravity equations by considering $\Phi_\text{bdy}\to+\infty$ and studying wavefunctions of $B_\text{ren}=B-2\log(\Phi_\text{bdy})$. The gravitational constraint for such asymptotic diamonds (and the relation with York time) was recently derived in \cite{Parrikar:2025xmz}, from the path integral. We show that $K_\text{grav}$ is consistent with their equation from the path integral in the asymptotic limit, \emph{after} conjugation by operators that arise from counterterms in the path integral. This supports our discussion for general diamonds in section \ref{sect:3.2quantumdiamonds}. To wrap up, in \textbf{section \ref{sect:3.4asym}} we briefly discuss asymmetric diamonds, where $\Phi$ takes unequal values $\Phi_\text{L}$ and $\Phi_\text{R}$ on the two diamond endpoints.

\subsection{Classical diamonds}\label{sect:3.1classicaldiamonds}
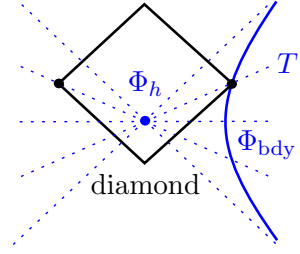
\begin{figure}
 \centering
 \begin{tikzpicture}[baseline={([yshift=-.5ex]current bounding box.center)}, scale=0.7]
 \pgftext{\includegraphics[scale=1]{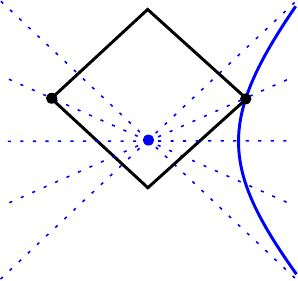}} at (0,0);
    \draw (-0,-1.2) node {diamond};
    \draw (-0,0.7) node {\color{blue}$\Phi_h$};
    \draw (2.3,-0.45) node {\color{blue}$\Phi_\text{bdy}$};
    \draw (2.7,1.1) node {\color{blue}$T$};
  \end{tikzpicture}
 \caption{Given $\Phi_\text{bdy}$, and using the boost isometry to put the two endpoints of the subregion at the same global time $\tau_\text{bdy}$, one completely fixes a diamond by specifying $(\Phi_h,T)$. Indeed, knowledge of $(\Phi_h,T,\Phi_\text{bdy})$ allows, using equation \eqref{3.1metric} and equation \eqref{3.2coord}, to determine the dilaton solution everywhere (by knowledge of $\Phi_h$), and the embedding of the diamond in the global spacetime (due to knowledge of $(\sigma_\text{bdy},\tau_\text{bdy})$). This differs from the space of subregions, which instead is fixed by specifying $(\Phi_h,T,s)$ given $\Phi_\text{bdy}$. See figure \ref{fig:diamond}.
 }
 \label{fig:subregions}
\end{figure}

We will consider two-sided black hole spacetimes in JT gravity, with metric and dilaton configurations:
\begin{equation}
    \d s^2=\frac{\d \sigma^2-\d \tau^2}{\cos(\sigma)^2}\,,\quad \Phi=\Phi_h\frac{\cos(\tau)}{\cos(\sigma)}\,,\quad -\frac{\pi}{2}<\sigma<\frac{\pi}{2}\,.\label{3.1metric}
\end{equation}
We are interested in studying the quantum description of subregions in this theory. We propose that one reasonable definition of a subregion is to fix the dilaton at the boundary of the subregion $\Phi\rvert_\text{bdy}=\Phi_\text{bdy}$. We now describe the space of subregions (modulo gauge transformations) with fixed boundary dilaton. It is convenient to consider the following coordinate transformation\footnote{Useful reference equations are (2.15) to (2.18) in \cite{Harlow:2018tqv}.}
\begin{equation}
    \tanh(\Phi_h T)=\frac{\sin(\tau)}{\sin(\sigma)}\,,\label{3.2coord}
\end{equation}
which brings the metric into the form: 
\begin{equation}
    \d s^2=-(\Phi^2-\Phi_h^2)\d T^2+\frac{\d \Phi^2}{\Phi^2-\Phi_h^2}\,.\label{3.3metric}
\end{equation}
One more coordinate transformation $\Phi=\Phi_h\cosh(\rho)$ brings the metric in the form $\d \rho^2-\Phi_h^2\cosh(\rho)^2\d T^2$. The subregion ends on two locations where $\Phi=\Phi_\text{bdy}$. This fixes one boundary coordinate $\cosh(\rho_\text{bdy})=\Phi_\text{bdy}/\Phi_h$. The diamond is then completely determined by the choice of time coordinates $T_1$ and $T_2$ in the two Rindler wedges (given that $\Phi_h$ was already chosen). Using the boost isometry of the spacetimes, we can choose these two boundary time coordinates to be $\abs{\text{equal}}$. Thus, a diamond with fixed $\Phi_\text{bdy}$ is uniquely fixed given $(\Phi_h,T)$. See figure \ref{fig:subregions}.

Alternatively and equivalently, one chooses the coordinate $\sigma_\text{bdy}$ and global time $\tau$ of the endpoints. The boundary coordinate $\sigma_\text{bdy}$ is directly related with the geodesic size $B$ of the subregion as follows:
\begin{equation}
    B=\int_{-\sigma_\text{bdy}}^{+\sigma_\text{bdy}}\frac{\d \sigma}{\cos(\sigma)}=2\,\text{arccosh}(1/\cos(\sigma_\text{bdy}))\,.\label{3.4B}
\end{equation}
Using \eqref{3.2coord} and \eqref{3.1metric} one can determine $\sigma_\text{bdy}(\Phi_h,T,\Phi_\text{bdy})$. Furthermore, using equation \eqref{3.4B}, one finds $\sinh(B/2)=\tan(\sigma_\text{bdy})$. Combining these one arrives at the following expression for the geodesic size $B$ of the subregions:
\begin{equation}
    \sinh(B/2)=\sqrt{\Phi_\text{bdy}^2/\Phi_\text{h}^2-1}\cosh(\Phi_h T)\,.\label{3.5B}
\end{equation}

\begin{figure}
 \centering
 \begin{tikzpicture}[baseline={([yshift=-.5ex]current bounding box.center)}, scale=0.7]
 \pgftext{\includegraphics[scale=1]{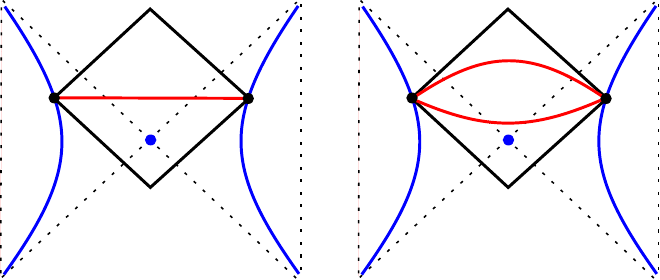}} at (0,0);
    \draw (-3,1.1) node {\color{red}$B$};
    \draw (3,0.6) node {\color{red}$s_1$};
    \draw (3,1.7) node {\color{red}$s_2$};
  \end{tikzpicture}
 \caption{The length $B$ of the geodesic slice of the diamond is determined given $(\Phi_h,T,\Phi_\text{bdy})$ via equation \eqref{3.5B}. Within any given diamond, slices with different extrinsic curvature or with different conformal time $s$ determine different initial data. The length $L$ and extrinsic curvature $K$ or a slice are determined as function of $(B,s)$ via equation \eqref{3.10phasespace}. Including $\Phi_h$, one completely determines also the dilaton profile on this slice (the subregion). The generator of changes in conformal time (a diffeomorphism) is a constraint.}
 \label{fig:diamond}
\end{figure}

The two endpoints of the subregions uniquely determine the causal diamond. Our goal is describing local gauge-invariant observables confined within those causal diamonds. To achieve this, we start with finding a convenient slicing of the causal diamond. For diamonds with endpoints at $\pm \sigma_\text{bdy}$ one can use the following conformal coordinates \cite{Jacobson:2018ahi,Banks:2020tox}:
\begin{equation}
    \d s^2=-\sin(\sigma_\text{bdy})^2\frac{-\d s^2+\d x^2}{(\cosh(s)+\cosh(x)\cos(\sigma_\text{bdy}))^2}\,,\quad -\infty<x<+\infty\,.\label{3.6metric}
\end{equation}
In analogy with our slicing of AdS closed universes in section \ref{sect:2.1setup}, fixed $s$ slices are slices with constant extrinsic curvature;
\begin{equation}
    K=\frac{\sinh(s)}{\sin(\sigma_\text{bdy})}\,.\label{3.7K}
\end{equation}
The lengths $L$ of these fixed curvature slices is computed as follows:
\begin{equation}
    L=\sin(\sigma^*)\int_{-\infty}^{+\infty}\frac{\d x}{\cosh(s)+\cosh(x)\cos(\sigma^*)}
\end{equation}
Using \eqref{3.7K}, this evaluates to
\begin{equation}
    \text{sinh}(\sqrt{1+K^2} L/2)=\sqrt{1+K^2}\tan(\sigma^*)
\end{equation}
Using furthermore $\sinh(B/2)=\tan(\sigma_\text{bdy})$ one obtains:
\begin{equation}
    \text{sinh}(\sqrt{1+K^2} L/2)=\sqrt{1+K^2}\sinh(B/2)\,,\quad K=\frac{\sinh(s)}{\tanh(B/2)}\,.\label{3.10phasespace}
\end{equation}
These equations determine $L(B,s)$ and $K(B,s)$. They are to be viewed as analogous to equations \eqref{2.4phasespace} and \eqref{2.7K} in section \ref{sect:2.1setup}. See figure \ref{fig:diamond}. In addition to the length $L$ and the curvature $K$ of our subregion, to fully fix the classical slice under investigation we are to specify the dilaton profile. As we see from equation \eqref{3.1metric}, for given choice of $\sigma_\text{bdy}$ (or equivalently $B$) different dilaton solutions are distinguished by the coordinate $\tau_\text{bdy}$ of the endpoints of the interval. One finds the following relation:
\begin{equation}
    \cos(\tau_\text{bdy})=\frac{\Phi_\text{bdy}/\Phi_h}{\cosh(B/2)}\,.\label{3.11tau}
\end{equation}
This allows to determine the full dilaton profile in the causal diamond, given knowledge of $(B,\Phi_h,\Phi_\text{bdy})$. We see for instance that lowering $\Phi_h$ (and keeping $B$ fixed) means that we are considering slices of the background \eqref{3.1metric} with lower $\tau$. This means the dilaton is closer to extremal. We note that the dilaton profile at the geodesic slice of the diamond is completely fixed by $(B,\Phi_\text{bdy})$:\footnote{The profile $\Phi\rvert_\text{geo}(x)$ is obtained by inserting the relation $\sigma(x,B)$, which one derives by relating the coordinate $x$ on the geodesic slice of the diamond \eqref{3.6metric} with the global $\sigma$ coordinate with metric $\d s=\d \sigma/\cos(\sigma)$.}
\begin{equation}
    \Phi\rvert_\text{geo}=\frac{\Phi_\text{bdy}}{\cosh(B/2)}\frac{1}{\cos(\sigma)}\,.\label{3.12phi}
\end{equation}
The dependence on $\Phi_h$ appears for instance in $\d \Phi/\d s$. In conclusion, the dilaton profile $\Phi(x)$ on a slice of interest is uniquely determined by $(B,s,\Phi_h)$, but slightly more complicated as compared to equation \eqref{2.6metric} for closed universes.

So, what is the phase space of subregions with fixed $\Phi_\text{bdy}$? Equivalently, what is the space of initial data on an interval with fixed $\Phi_\text{bdy}$? We restrict to minisuperspace, by only considering configurations with uniform extrinsic curvature $K$. We expect that such a restriction is exact (using spatial diffeos to restrict to uniform configurations). It would be interesting to derive this. We explained that there is a three dimensional space of such initial data labeled by $(B,\Phi_h,s)$ - in complete analogy with the closed universe case in section \ref{sect:2.1setup}. An equivalent parameterization is $(T,\Phi_h,s)$.

The more unusual coordinate to include here is $s$, representing the conformal time of slices. Because $s$ represents a coordinate transformation, in gravity $s$ should be a redundant variable. We appreciate this because different choices of $s$ represent different slices of the same geometry (with identical endpoints). Nevertheless, for the purpose of constructing gauge-invariant local matter operators in section \ref{sect:4obssubr}, we find it convenient to retain $s$ as phase space variable. This means we should also retain the generator of $s$ transformations, which is a CKV flow \cite{Jacobson:2018ahi}. We call this generator, which is a gravitational constraint, $K_\text{grav}$.

In section \ref{sect:3.2quantumdiamonds} we discuss the quantization of this 4d phase space. Harlow and Jafferis \cite{Harlow:2018tqv} discussed the quantization of the physical phase space $K_\text{grav}=0$, for the case of asymptotic diamonds $\Phi_\text{bdy}\to\infty$. We consider asymptotic diamonds in section \ref{sect:3.3assdiamond}.

Thus far we silently assumed $\Phi_\text{bdy}>\Phi_h$. In the discussion section \ref{sect:concl} we briefly consider $\Phi_\text{bdy}>\Phi_h$, in which case the diamonds lie completely inside the black hole interior. We will furthermore comment on an alternative definition of diamonds in quantum gravity, where one fixes instead $p\rvert_\text{bdy}=p_\text{bdy}$.

\subsection{Quantum diamonds and towards interior physics from finite cutoff quantization}\label{sect:3.2quantumdiamonds}
We next consider the Hilbert space of diamonds (or subregions). We claim that the symplectic form is identical to equation \eqref{2.25poisson}:
\begin{equation}
    \omega=\d K_\text{grav}\wedge \d S+\d B\wedge \d P\,.\label{3.13omega}
\end{equation}
Let us explain the different ingredients in this equation, and motivate why this equation is correct. In the case of assymptotics diamonds $\Phi_\text{bdy}\to\infty$, we explain in section \ref{sect:3.3assdiamond} that this equation follows from the gravitational path integral. It would be interesting to have a first principles path integral derivation of \eqref{3.13omega}.

The phase space variable $S$ represents a choice of (conformal) time slice within the diamond \eqref{3.6metric}. As shown in \cite{Jacobson:2018ahi}, $S$ represents a CKV flow. This fact (together with the realization that $S$ flow equals modular flow in the global vacuum state \cite{Jacobson:2018ahi,Casini:2011kv}) allows for a straightforward dressing of local conformal matter correlators to ``geometric time'' $S$ in section \ref{sect:4obssubr} - mimicking the construction of section \ref{sect:2obcl}. $K_\text{grav}$ is the generator of $S$ transformations, therefore the symplectic form includes the combination $\d K_\text{grav}\wedge \d S$. As in equation \eqref{2.30constraint} this generator is a certain (integrated) component of the gravitational stress tensor a.k.a. Einstein tensor:
\begin{equation}
    K_\text{grav}=\int_{-\infty}^{+\infty}\d x\,\sqrt{h}n^\mu \xi^\nu T_{\text{grav}\,\mu\nu}\,,\quad T_{\text{grav}\,\mu \nu}=  \nabla_\mu\nabla_\nu \Phi-g_{\mu\nu}\Box \Phi+g_{\mu\nu}\Phi \,.\label{3.14k}
\end{equation}
Explicitly, using the metric \eqref{3.6metric} of the causal diamond and introducing (following the notation of \cite{Jacobson:2018ahi}, appendix B)
\begin{equation}
    C=\frac{\cosh(s)+\cosh(x)\cos(\sigma_\text{bdy})}{\sin(\sigma_\text{bdy})}\,\label{3.15C},
\end{equation}
one obtains
\begin{equation}
    \sqrt{h}n^\mu \xi^\nu T_{\text{grav}\,\mu\nu}=\bigg \lbrace \frac{\d^2}{\d x^2}+\frac{1}{C}\frac{\d C}{\d x}\frac{\d}{\d x}+\frac{1}{C}\frac{\d C}{\d s}\frac{\d}{\d s}-\frac{1}{C^2}  \bigg \rbrace \Phi\,.\label{3.16C}
\end{equation}
This should be compared with equation \eqref{2.58tgrav}. We remark that for asymptotic diamonds (with $\sigma_\text{bdy}=\pi/2$) that this equation exactly reduces to the closed universe equation \eqref{2.58tgrav}. See section \ref{sect:3.3assdiamond}.

We would now like to argue that the remainder of the symplectic form is $\d B\wedge \d P$. Here, we defined $P$ as the canonical conjugate of $B$. The relevant point is arguing for this simple dependence on $B$. This form of the symplectic form requires that $B$ is ``orthogonal'' to $K_\text{grav}$:
\begin{equation}
    [B,\textbf{K}_\text{grav}]=0\,.
\end{equation}
Because $K_\text{grav}$ is a gravitational constraint, this means that $B$ is gauge-invariant, or more precisely that changing $S$ does not alter $B$ (this also follows from the Hamilton equations with Hamiltonian $N K_\text{grav}$). This is obvious from our definition of $B$! Indeed, in the metric \eqref{3.6metric} the coordinate $x$ does not change with $s$, neither does $\sigma_\text{bdy}$. This means that the following function of $(x_1,x_2,\sigma_\text{bdy})$ commutes with $K_\text{grav}$
\begin{equation}
    B(x_1,x_2)=\sin(\sigma_\text{bdy})\int_{x_1}^{x_2}\frac{\d x}{1+\cosh(x)\cos(\sigma_\text{bdy})}\,.
\end{equation}
The special case of the geodesic length of the slice hence also commutes with $K_\text{grav}$! For closed universes, this is equivalent to the statement that the quantities in equation \eqref{2.24bp} commute with \eqref{2.20hgrav}. More in general: one can always split up the symplectic form into a piece involving the constraint, and a piece involving invariants that commutes with the constraint $K_\text{grav}$:
\begin{equation}
    \omega=\d K_\text{grav}\wedge \d S + \omega_\text{invariant}\,.
\end{equation}
An explicit example for closed universes is equation \eqref{2.24bp}. All we have done in \eqref{3.13omega} is claim that the geodesic size $B$ of the subregion is $s$ invariant (which is obvious), and chosen it as one of the coordinates on the invariant phase space (this choice we are free to make)\footnote{An identical choice of gauge-invariant variable seems to generalize well to higher dimensions. As advocated in \cite{witten2023notecanonicalformalismgravity} (and references therein), gauge-fixing to maximal volume slice works well for perturbative gravity in higher dimensional AdS.}. It is the quantization of this invariant space that was discussed for instance by Harlow-Jafferis \cite{Harlow:2018tqv}.\footnote{Closely related perspectives were presented in \cite{Blommaert:2018oro,Yang:2018gdb,Iliesiu:2019xuh,Saad:2019pqd}.} Let us briefly clarify this relation.

As in section \ref{sect:2.2Ktime} states are described by wavefunctions $\psi(B,S)$. Physical states satisfy $\mathbf{K}_\text{grav}\ket{\psi}=0$ and therefore have $S$-independent wavefunctions $\psi(B,S)=\psi(B)$ \cite{Held:2024rmg}. Another natural set of physical coordinates on phase space are $(E_\text{Rindler},T)$ with the following symplectic form:
\begin{equation}
    \omega_\text{invariant}=\d B\wedge \d P=\Phi_h\d \Phi_h\wedge \d T=\d E_\text{Rindler}\wedge \d T\,,\label{3.20omega}
\end{equation}
with $B(\Phi_h,T)$ given in equation \eqref{3.5B}. To show that this symplectic form is correct we perform another change of coordinates to proper time in the metric \eqref{3.3metric}. This results in
\begin{equation}
    \omega_\text{invariant}=\d E_\text{ADM}\wedge \d T_\text{proper}\,,\quad E_\text{ADM}=-\sqrt{\Phi_\text{bdy}^2-\Phi_h^2}\,.\label{3.21}
\end{equation}
This expression for the conjugate $E_\text{ADM}$ of boundary proper time is indeed found, for instance, in \cite{Iliesiu:2020zld}.\footnote{One can essentially derive this from the equations in section \ref{sect:2obcl}. Indeed boundary length $\ell$ is conjugate to $p=\sqrt{\Phi_h^2-\Phi^2}$. Then we transform to Lorentzian length changes.} To relate with the discussion of Harlow-Jafferis \cite{Harlow:2018tqv}, we briefly show how wavefunctions and amplitudes of finite cutoff JT gravity ($\Phi_\text{bdy}$ is finite) arise \cite{Iliesiu:2020zld,Griguolo:2025kpi,Griguolo:2021wgy,Stanford:2020qhm}. Instead of studying physical wavefunctions of $B$, we may study for instance physical wavefunctions $\psi(E_\text{Rindler})$. Their overlaps are found by solving (as differential equation acting on wavefunctions $\psi(B)$)
\begin{equation}
    E_\text{Rindler}=P^2+\frac{\Phi_\text{bdy}^2}{\cosh(B/2)^2}\,,\label{3.22de}
\end{equation}
which one finds by inverting (using the symplectic form) equation \eqref{3.5B}, which reads in terms of $E_\text{Rindler}$
\begin{equation}
    \sinh(B/2)=\sqrt{\Phi_\text{bdy}^2/E_\text{Rindler}-1}\cosh(\sqrt{E_\text{Rindler}}T_\text{Rindler})\,.
\end{equation}
One could then for instance compute the disk amplitude using canonical quantization as the expectation value of $e^{L \sqrt{\Phi_\text{bdy}^2-\mathbf{E}_\text{Rindler}}}$ in the (tracial) state $\ket{\mathbf{B}=0}$, by inserting completion relations in $E_\text{Rindler}$ basis. Similarly, one could compute the two-point matter correlation functions using expectation values of $e^{L_1 \sqrt{\Phi_\text{bdy}^2-\mathbf{E}_\text{Rindler}}}\,G_{m^2}(\mathbf{B})\,e^{L_2 \sqrt{\Phi_\text{bdy}^2-\mathbf{E}_\text{Rindler}}}$ where $G_{m^2}(d)$ is the bulk-bulk propagator of a massive scalar which depends on geodesic distance $d$ between the two insertion points. Note indeed that because $\Phi_\text{bdy}$ remains finite that the matter correlators are not evaluated in an asymptotic limit. In the asymptotic limit, to which we turn next, these wavefunctions become the Bessel functions of Harlow-Jafferis \cite{Harlow:2018tqv} and one recovers the usual JT formulas. We understand that an evaluation of these quantum mechanical amplitudes has currently been investigated \cite{Jacopowip}. We remark that \eqref{3.22de} has solutions also for $\Phi_h>\Phi_\text{bdy}$, which formally suggests that the boundary points of the diamonds are in the interior of \eqref{3.2coord}. Indeed the potential \eqref{3.22de} suggests that, in this case, real solutions for $B$ go through zero (and may formally flip sign between black hole and white hole interiors). So, this quantum mechanics may describe physics in black hole interiors. It would clearly be interesting to understand this QM in detail. We thank Adam Levine, Jacopo Papalini and Aron Wall for discussions on this. We will comment briefly on this in the discussion (section \ref{sect:concl}).

We note that the finite cutoff JT path integral provides a way of preparing ``natural'' states for our subregion algebra (including matter, which we will discuss in section \ref{sect:4obssubr}) - by inserting matter operators on the boundary with finite $\Phi_\text{bdy}$, mimicking similar preparations for asymptotic JT gravity in \cite{Kolchmeyer:2023gwa,Penington:2023dql}.

Finally, note that equation \eqref{3.10phasespace} giving $L(B,S)$ and $K(B,S)$ may be used to write the gravitational constraint in terms of the momenta $(P,-Q)$ conjugate to $(L,K)$:\footnote{It is important that we are doing a simple coordinate change on phase space that does not directly involve the conjugate momenta. A transformation $\tilde{q}(q,p)$ leads to complicated differential equations for the conjugate momenta $\tilde{p}$. However, for $\tilde{q}(q)$ one indeed simply computes $\tilde{p}$ using the chain rule.}
\begin{equation}
    K_\text{grav}=\frac{\d K}{\d S}(L,K)\,Q-\frac{\d L}{\d S}(L,K)\,P\,.\label{3.24}
\end{equation}
For closed universes this reproduces the WDW gravitational constraint \eqref{2.20hgrav} as deduced from the path integral. We now show that in the case of asymptotic diamonds ($\Phi_\text{bdy}\to\infty$), indeed this equation also reproduces the formula for the WDW constraint as derived by \cite{Parrikar:2025xmz} using the gravitational path integral. This is additional evidence that equation \eqref{3.24} is correct for generic diamonds.

\subsection{Asymptotic diamonds}\label{sect:3.3assdiamond}
\begin{figure}
 \centering
 \begin{tikzpicture}[baseline={([yshift=-.5ex]current bounding box.center)}, scale=0.7]
 \pgftext{\includegraphics[scale=1]{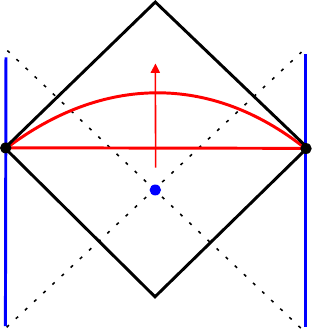}} at (0,0);
    \draw (-0,-2.5) node {diamond};
    \draw (-4.1,-0.5) node {\color{blue}asymptotic};
    \draw (-4.1,-1.2) node {\color{blue}boundary};
    \draw (-0,2) node {\color{red}time};
    \draw (3.85,-0.5) node {\color{blue}$\Phi_\text{bdy}=\infty$};
  \end{tikzpicture}
 \caption{Asymptotic diamonds (where $\Phi_\text{bdy}=\infty$) have their endpoints on the usual holographic boundary $\sigma_\text{bdy}=\pi/2$. The geodesic length $B$ of the diamond diverges, but one can describe meaningful (quantum) mechanics using a renormalized length $B_\text{ren}$. The space of subregions is described by $(B_\text{ren},\Phi_h,s)$, for given $\Phi_\text{bdy}$.}
 \label{fig:assdiamond}
\end{figure}
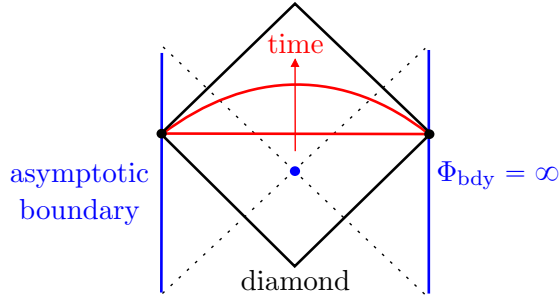

We next discuss the special case of asymptotic diamonds, where equations simplify and one can compare with more conventional JT equations. For asymptotic diamonds one takes $\Phi_\text{bdy}\to\infty$, which according to equation \eqref{3.1metric} implies $\sigma_\text{bdy}=\pi/2$. We have pictured those diamonds in figure \ref{fig:assdiamond}. In this regime, the metric and dilaton solutions \eqref{3.6metric} and \eqref{3.1metric} become
\begin{equation}
    \d s^2= \frac{-\d s^2+\d x^2}{\cosh(s)^2}\,,\quad \Phi=\Phi_h\frac{\cosh(x)+\sinh(s)\sinh(s_h)}{\cosh(s)\cosh(s_h)}\,.\label{3.25metric}
\end{equation}
Because the dilaton solutions are significantly simpler than in the general $\Phi_\text{bdy}$ case we have given them explicitly. The variable $s_h$ labels the location of the saddle of the dilaton. It is related to $\tau_\text{bdy}$ through
\begin{equation}
    \cosh(s_h)=\frac{1}{\cos(\tau_\text{bdy})}\,.\label{3.26sh}
\end{equation}

Clearly, in the metric \eqref{3.25metric}, the geodesic size $B$ of the diamond diverges. However, as we see from equation \eqref{3.5B} one can define a renormalized version $B_\text{ren}$ of the geodesic size of the subregion that \emph{does} remain finite for $\Phi_\text{bdy}\to\infty$, as follows:
\begin{equation}
    B=2\log 2\Phi_\text{Bdy}+B_\text{ren}\,,\quad B_\text{ren}=2\log \cosh(\Phi_h T)\,.\label{3.27bren}
\end{equation}
The Rindler energy eigenvalue equation \eqref{3.22de} then simplifies to the famous Liouville equation describing the evolution of the size of (asymptotic) two-sized slices in JT gravity \cite{Harlow:2018tqv}:
\begin{equation}
    E_\text{Rindler}=P_\text{ren}^2+e^{-B_\text{ren}}\,.\label{3.28erin}
\end{equation}
The phase space variables \eqref{3.10phasespace} become
\begin{equation}
    \frac{1}{2\Phi_\text{bdy} \sqrt{1+K^2}}e^{\sqrt{1+K^2}L/2}=e^{B_\text{ren}/2}\,,\quad K=\sinh(s)\,.\label{3.29bren}
\end{equation}
Via \eqref{3.27bren} these are expressed as function of the three physical phase space variables $(T,\Phi_h,s)$ for fixed $\Phi_\text{bdy}$. The dilaton solution \eqref{3.25metric} is expressed in terms of $(T,\Phi_h,s)$ by using the renormalized version of equation \eqref{3.11tau} which combined with equation \eqref{3.26sh} reads
\begin{equation}
    \cosh(s_h)=\Phi_h\cosh(\Phi_h T)\,.
\end{equation}
The inner product on physical states is the usual one on square integrables of $B_\text{ren}$ (which ranges from $-\infty$ to $+\infty$), and the Rindler energy wavefunctions are the Bessel functions of Harlow and Jafferis \cite{Harlow:2018tqv}.\footnote{Amusingly this inner product can also be derived from group averaging using the WDW Hamiltonian $H=\i \hbar \d/\d T$ with $T$ proper time on (the observer's worldline $x=0$), unifying the inner products for open and closed universes in JT gravity. We leave the derivation as an exercise for the interested reader \cite{Held:2024rmg}. The inner product derived from group averaging has a $B$ integration measure, leading to $P^2\to B^{-1/2}P^2 B^{1/2}$ in \eqref{3.28erin} as Hermitian operator with respect to this inner product. This conjugation by $B^{1/2}$ otherwise does not affect the Hilbert space or amplitudes. The $B$ measure factor arises just like in the closed universe group-averaging calculation \cite{Held:2024rmg}.} As we explained at length, $B$ is a gauge-invariant variable, and according to equation \eqref{3.27bren} therefore so is $B_\text{ren}$. This means that physical wavefunctions $\psi(L,K)$ only depend on the combination $B_\text{ren}(L,K)$:
\begin{equation}
    \psi(L,K)=\psi(\sqrt{1+K^2}L-2\log 2\Phi_\text{bdy}\sqrt{1+K^2})\,.
\end{equation}
This is consistent with equations in \cite{Yang:2018gdb}, derived from the gravitational path integral: all gravitational wavefunctions only depend on this specific combination $B_\text{ren}$. See for instance also equation (4.51) and equation (4.52) in \cite{Parrikar:2025xmz}.

This should be considered a consequence of the gravitational constraint \eqref{3.24}, which for asymptotic diamonds evaluates to:
\begin{equation}
    K_\text{grav}=\sqrt{1+K^2}\,Q+\bigg( \frac{KL}{\sqrt{1+K^2}}-\frac{2K}{1+K^2} \bigg) P\,.\label{3.32kgrav}
\end{equation}
We would now like to compare with a calculation of the ``York Hamiltonian'' for asymptotic diamonds from the gravitational path integral, derived in appendix C of \cite{Parrikar:2025xmz}. See their equation (4.79). The York constraint is the generator of changes in extrinsic curvature $\sinh(s)$. This is related to the generator of $s$ translations $K_\text{grav}$ \eqref{3.32kgrav} by the Jacobian $1/\cosh(s)=1/\sqrt{1+K^2}$:\footnote{One could interpret this equation as arising due to changing polarization from $\psi(K,L)$ (the right hand side) to $\psi(K,B)$ (the left hand side, instead of previously $\psi(S.B)$. The transformation $L\to L(K,B)$ then indeed results in this chain rule.}
\begin{equation}
    H_\text{York}=Q+\frac{KL}{1+K^2} P +\,\text{``subleading''.}\label{3.33}
\end{equation}
One might argue that the last term is ``subleading'' in the asymptotic limit, because in the expression $\sqrt{1+K^2}L-2\log 2\Phi_\text{bdy}\sqrt{1+K^2}$ the term $2\log\sqrt{1+K^2}$ does not grow with $\Phi_\text{bdy}$, and the $K$ derivative of that term is responsible for the ``subleading'' contribution in \eqref{3.33}. Such terms were not captured by the derivation in appendix C of \cite{Parrikar:2025xmz}, but we believe that they should be there. It would be interesting to prove this explicitly. To complete the relation with the computation of \cite{Parrikar:2025xmz}, we perform a similarity transformation. The need for this similarity transformation is that in the computation of gravitational wavefunctions using the path integral in \cite{Yang:2018gdb,Parrikar:2025xmz}, certain counterterms were added to the action. These modify the wavefunctions:
\begin{equation}
    \psi_\text{path\,integral}(K,L)=e^{-2\i\Phi_\text{bdy}\text{arcsinh}(K)}\psi(K,L)\,.
\end{equation}
The gravitational path integral with counterterms then satisfies a ``conjugated'' constraint
\begin{equation}
    H_\text{York\,path\,integral}=e^{-2\i\Phi_\text{bdy}\text{arcsinh}(K)} H_\text{York}\, e^{+2\i\Phi_\text{bdy}\text{arcsinh}(K)}=Q+\frac{KL}{1+K^2} P +\frac{2\Phi_\text{bdy}}{\sqrt{1+K^2}}+\,\text{``subleading''.}\label{3.35hyork}
\end{equation}
This indeed reproduces equation (4.79) in \cite{Parrikar:2025xmz}, demonstrating in an explicit example that our constraint \eqref{3.32kgrav} arises from the gravitational path integral. We note that Harlow-Jafferis type equations such as the Liouville Hamiltonian only nicely arise in the un-conjugated formulation. Indeed, equation \eqref{3.28erin} does not commute with \eqref{3.35hyork}. We note that demanding commutation with the (clearly gauge-invariant) renormalized length \eqref{3.29bren} uniquely fixed the ``subleading'' terms in the constraint \eqref{3.33} to \eqref{3.32kgrav}.

\subsection{Asymmetric diamonds}\label{sect:3.4asym}
\begin{figure}
    \centering
    \begin{tikzpicture}[baseline={([yshift=-.5ex]current bounding box.center)}, scale=0.7]
 \pgftext{\includegraphics[scale=1]{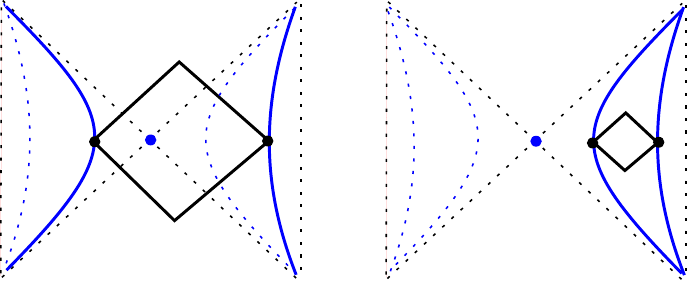}} at (0,0);
    \draw (-4.9,0.8) node {\color{blue}$\Phi_\text{L}$};
    \draw (4.35,1.5) node {\color{blue}$\Phi_\text{L}$};
    \draw (5.975,-0.9) node {\color{blue}$\Phi_\text{R}$};
  \end{tikzpicture}
    \caption{Given the boundary dilatons and global time, there's still a two-fold ambiguity about the subregion: whether the two end points are on the same, or the opposite sides of the bifurcation surface. One diamond (left) has $\sigma_\text{L}<0$ whereas the other (right) has $\sigma_\text{L}>0$.}
    \label{fig:twosubregions}
\end{figure}

As a generalization of the previous calculations we consider the case where the diamond is asymmetric. With this we mean that the dilaton value at the two endpoints of the diamond differs. Let us consider the global metric and dilaton as given in (\ref{3.1metric}). Suppose the dilaton values at the endpoints are $\Phi_\text{L}$ and $\Phi_\text{R}$. Given the global time $\tau_{\rm bdy}$ (which we may choose equal for both endpoints), the spatial locations of the endpoints are determined up to a sign ambiguity $\sigma_\text{L}\to-\sigma_\text{L}$:
\begin{align}
    \cos(\sigma_\text{R}) =\frac{\Phi_h}{\Phi_\text{R}}\cos(\tau_{\rm bdy})\,,\quad \cos(\sigma_\text{L}) =\frac{\Phi_h}{\Phi_\text{L}}\cos(\tau_{\rm bdy})\,,
\end{align} 
See Figure \ref{fig:twosubregions}. Basically, the two end points can be on either the same or opposite sides of the bifurcation surface. In the formula below, we will use the spatial locations $\sigma_R$ and $\sigma_L$ directly, hence the ambiguity will not bother us. 

We now find the diamond conformal coordinates as function of $(\sigma_\text{L},\sigma_\text{R})$. Without loss of generality, we may perform these calculations for $\tau_{\rm bdy}=0$. As in the symmetric case, we introduce null coordinates: $u=\tau+\sigma,$ $v=\tau-\sigma,$ and then go to the``diamond null coordinates'' $(x^+,x^-)$:
\begin{align}
    e^{x^+} = \frac{\sin \frac{u-u_\text{L}}{2}}{\sin\frac{u_\text{R}-u}{2}}\,,\quad e^{x^-} = \frac{\sin \frac{v-v_\text{L}}{2}}{\sin\frac{v_\text{R}-v}{2}}\,.
\end{align}
Introducing $s=(x^++x_-)/2$ and $x=(x^+-x^-)/2$ the conformal metric in the asymmetric diamond is:
\begin{align}\label{eq:asymmetricdiamond}
    \d s^2 = \frac{-\d s^2+\d x^2}{C^2}\,,\quad C = \frac{\cos\frac{\sigma_\text{R}+\sigma_\text{L}}{2}\cosh(s) + \frac{1}{2} e^x\cos(\sigma_\text{R})+\frac{1}{2}e^{-x}\cos(\sigma_\text{L})}{\sin \frac{\sigma_\text{R}-\sigma_\text{L}}{2}}\,.
\end{align} 
This indeed reduces to \eqref{3.15C} when $\sigma_\text{L}=-\sigma_\text{R}$. This may be written in a more gauge-invariant manner, by writing $\sigma_\text{L}(B,\Phi_\text{L},\Phi_\text{R})$ and $\sigma_\text{R}(B,\Phi_\text{L},\Phi_\text{R})$ using (this equation assumes that $\sigma_\text{R}>0$ and $\sigma_\text{R}>\abs{\sigma_\text{L}}$)
\begin{equation}
    B=\int_{\sigma_\text{L}}^{\sigma_\text{R}}\frac{\d \sigma}{\cos(\sigma)}=\,\text{arccosh}(1/\cos(\sigma_\text{R}))-\text{sgn}(\sigma_\text{L})\,\text{arccosh}(1/\cos(\sigma_\text{L}))\,,\quad \frac{\Phi_\text{R}}{\Phi_\text{L}}=\frac{\cos(\sigma_\text{L})}{\cos(\sigma_\text{R})}\,.
\end{equation}
Furthermore the constant curvature of fixed conformal time slices is
\begin{align}
    K= \frac{\cos\frac{\sigma_\text{R}+\sigma_\text{L}}{2}}{\sin\frac{\sigma_\text{R}-\sigma_\text{L}}{2}}\sinh(s)\,.
\end{align}
The conformal isometry is of course $\xi=\d/\d s$. It would be interesting to know the analogue of the ADM respectively Rindler energy in this setup, generating respectively boundary proper time or Rindler time evolution. It seems natural to introduce either the right-or left ADM (or Rindler) energy as phase space variables. The sum of left and right ADM energies may also make sense. We leave this for future work.

Having concluded our discussion of the phase space and canonical quantization of subregions in pure JT gravity, we next include conformal matter. Following in close analogy the construction of section \ref{sect:2obcl} (for closed universes), we will construct gauge-invariant local conformal matter correlators, and discuss the associated type II subregion gravitational entropy.

\section{Observables in gravitational subregions}\label{sect:4obssubr}
In this section we construct gauge invariant local conformal matter observables in quantum subregions, by combining (in a relatively straightforward way) the ingredients of section \ref{sect:2obcl} and section \ref{sect:3diamons}. In \textbf{section \ref{sect:4.1clocks}}, we recall \cite{Casini:2011kv,Jacobson:2018ahi} that the modular Hamiltonian of the global CFT vacuum state, restricted to our subregions, generates conformal time translations. This leads to a simple expression $K_\text{grav}+K_\text{matter}=0$ for the gravitational constraint associated with CKV flow which we leverage in \textbf{section \ref{sect:4.2relational}} to construct gauge invariant local observables. The main extra ingredient as compared to our calculations of section \ref{sect:2obcl} is the check that including matter stress energy does not affect the structure of the gravitational part of the Hilbert space. This happens also for closed universes \cite{Held:2024rmg}, because the Einstein equations \eqref{2.26eeq} are linear equations for the dilaton $\Phi$. In \textbf{section \ref{sect:4.3entropy}}, we compute the gravitational entropy of certain semiclassical states. The entropy is associated with the type II$_\infty$ algebra of gauge-invariant observables (conform section \ref{sect:2.4closedentropy}). We find that this is a \emph{bulk} quantity. For reference, we check (or, recall) that an identical calculation in the case of a Rindler wedge leads to the usual Bekenstein-Hawking area formula.

\subsection{Curvature clocks revisited}\label{sect:4.1clocks}
Recall that our diamonds are defined using the coordinates of the endpoints $(\sigma_\text{bdy},\tau_\text{bdy})$ of the diamond using the metric and dilaton solution in global coordinates \eqref{3.1metric}:
\begin{equation}
    \d s^2=\frac{\d \sigma^2-\d \tau^2}{\cos(\sigma)^2}\,,\quad \Phi=\Phi_h\frac{\cos(\tau)}{\cos(\sigma)}\,,\quad -\frac{\pi}{2}<\sigma<\frac{\pi}{2}\,.
\end{equation}
We now include conformal matter on this spacetime, prepared in the global (meaning $\tau$) vacuum state using the usual Euclidean path integral preparation. In the diamond, we will use the coordinates \eqref{3.6metric}:
\begin{equation}
    \d s^2=-\sin(\sigma_\text{bdy})^2\frac{-\d s^2+\d x^2}{(\cosh(s)+\cosh(x)\cos(\sigma_\text{bdy}))^2}\,,\quad -\infty<x<+\infty\,.\label{4.2metric}
\end{equation}
The question we want to answer here is: given that the conformal matter state is in the global vacuum, how is this perceived from \emph{within} the subregion, using the coordinate $s$ as the time? This question was answered by Casini-Huerta-Meyers \cite{Casini:2011kv} and Jacobsen-Visser \cite{Jacobson:2018ahi}. \emph{If}, as in equation \eqref{2.11rescale}, we consider rescaled conformal matter operators that absorb the conformal transformation between the metric \eqref{4.2metric} and the flat metric (these factors are thus a function of the function $C$ defined in \eqref{3.16C}), \emph{then} one finds that the matter state on the subregion is a thermal state with respect to $s$ evolution \eqref{2.15kms}:
\begin{equation}
    \rho_\text{matter}=e^{-2\pi K_\text{matter}/\hbar }\,,\quad [a(s),K_\text{matter}]=\i\hbar \frac{\d}{\d s}a(s)\,.\label{4.3kmat}
\end{equation}
In other words, the modular Hamiltonian $-\log \rho_\text{matter}$ determined by the global vacuum is (proportional to) the generator of conformal time translations $K_\text{matter}$. Conformal matter is thermal with respect to conformal time (or CKV) flow! General matter is thermal with respect to modular flow (by definition), but this does not buy one much in terms of constructing gauge-invariant local operators, unless if the modular flow is geometric \cite{Sorce:2024zme,Mertens:2025rpa}. For conformal matter, the known point which we are re-emphasizing here is simply that indeed modular flow for our subregions in AdS \emph{is} geometric, and equal to $s\to s+c$.

Since we defined $K_\text{matter}$ as the generator of $s\to s+c$ in \eqref{4.3kmat}, the associated gravitational constraint takes a simple form. More precisely, projecting Einstein's equations \eqref{2.26eeq} using the coordinates \eqref{4.2metric} onto the component associated with the generator of conformal time flow using
\begin{equation}
    K_\text{matter}=\int_{-\infty}^{+\infty}\d x\,\sqrt{h}n^\nu\xi^\nu T_{\mu \nu}\,,
\end{equation}
one finds that the relevant gravitational constraint becomes the analogue of equation \eqref{2.30constraint} (in the case of closed universes):
\begin{equation}
    K_\text{grav}+K_\text{matter}=0\,.\label{4.5con}
\end{equation}
Recall that $K_\text{grav}$ was discussed in \eqref{3.24} and is the conjugate to the phase space variable $S$ (see equation \eqref{3.13omega}). The complete analogy with the closed universe discussion of section \ref{sect:2.2Ktime} is now hopefully clear.

\subsection{Relational observables revisited}\label{sect:4.2relational}
We now consider the gauge-invariant operator algebra. If the equivalence with the structure in the case of closed universes were complete, then we may copy the conclusions of section \ref{sect:2.3closedobs}. In particular, gauge invariant observables consist of functions $f(\mathbf{K}_\text{grav})$, square integrables of $\mathbf{B}$ and dressed local conformal matter operators (the physically most interesting class):
\begin{equation}
    a(s+\mathbf{S})=e^{\i K_\text{matter}\mathbf{S}/\hbar}\,a(s)\,e^{-\i K_\text{matter}\mathbf{S}/\hbar}\,.
\end{equation}
The matter algebra is type III (because we are considering a subregion in QFT), but as in CLPW \cite{Chandrasekaran:2022cip}, the gauge-invariant algebra is type II$_\infty$. The associated trace is identical to \eqref{2.44trace}, and entropy differences of semiclassical gravitational states (keeping the matter state fixed) are computed as in equation \eqref{2.54entropyv1}:
\begin{equation}
    \delta S=-\frac{2\pi}{\hbar}\delta K_\text{grav}\,.\label{4.7S}
\end{equation}
To make the analogy with the discussion of section \ref{sect:2.3closedobs} complete, and thus prove the validity of the above statements and equations, we require one new ingredient. We are to show that the gravitational sector and the matter sector of the Hilbert space \emph{decouple} prior to imposing the constraint \eqref{4.5con}. Otherwise, if matter would ``backreact'' on the gravitational sector, then the proof that \eqref{2.44trace} is tracial no longer would hold. In the remainder of this section we show that this decoupling does occur. This statement is well known for closed universes \cite{Held:2024rmg}, and for asymptotic diamonds \cite{Penington:2023dql}.

All we need to show is that, in the presence of matter, the dilaton solution of the Einstein equation \eqref{2.26eeq} decomposes as
\begin{equation}
    \Phi=\Phi_\text{bare}+\Phi_\text{sourced}\,,
\end{equation}
with $\Phi_\text{bare}$ the (previously discussed) vacuum solution, and with the effects of the matter perturbations captured by the particular solution: $\Phi_\text{sourced}$. Since both $\Phi$ and $\Phi_\text{bare}$ have (by construction) boundary value $\Phi_\text{bdy}$, we require that the particular solution vanishes at the boundary:
\begin{equation}
    \Phi_\text{sourced}\rvert_{\text{bdy}}=0\,.\label{4.9sourced}
\end{equation}
If this solution exists and is unique, then the phase space remains determined by $\Phi_\text{bare}$ and we would be in the scenario where the gravitational sector and the matter sector of the Hilbert space do decouple. What could go wrong is that there may be no sourced solution with homogeneous boundary conditions \eqref{4.9sourced}. In that case, $\Phi_\text{bare}$ and the gravitational part of phase space would depend on the matter content, so there would be backreaction. We now demonstrate (by explicit construction) that a unique solution to the sourced equations with boundary conditions \eqref{4.9sourced} does exist, thus proving the claimed structure and in particular the entropy formula \eqref{4.7S}

We first consider asymptotic diamonds with metric \eqref{3.25metric}. We seek a solution to the matter-sourced Einstein's equations \eqref{2.26eeq}. We will consider a stationary massive particle with energy $E$ at $x=0$ - for which the stress energy only has a time-time component:
\begin{equation}
    \bigg\{\cosh(s)\frac{\d^2}{\d x^2}+\sinh(s)\frac{\d}{\d s}-\frac{1}{\cosh(s)} \bigg\}\Phi_\text{sourced}=-H_\text{matter}=-E\,\delta(x)\,.\label{4.10source}
\end{equation}
The unique solution is found by patching two vacuum solutions together with suitable jump conditions at $x=0$:
\begin{equation}
    \frac{\d }{\d x}\Phi_\text{sourced}\rvert_\varepsilon-\frac{\d }{\d x}\Phi_\text{sourced}\rvert_{-\varepsilon}=-\frac{E}{\cosh(s)}\,.
\end{equation}
The general vacuum solution is \cite{Gao:2021uro}
\begin{equation}
    \Phi=\frac{\alpha \cosh(x+\beta)+\gamma\sinh(s)}{\cosh(s)}\,.
\end{equation}
The unique solution follows by additionally imposing the boundary conditions \eqref{4.9sourced} $\Phi_{\text{sourced}}\rvert_{x\to\pm\infty}=0$:
\begin{equation}
    \Phi_\text{sourced}=\frac{E}{2}\frac{e^{-\abs{x}}}{\cosh(s)}\,.\label{4.13phisource}
\end{equation}
The combined solution $\Phi_\text{bare}+\Phi_\text{sourced}$ has minima at a nonzero value of $\abs{x}$. So, the spatial wormhole has become larger. This is the usual effect of adding positive matter density in AdS. How about finite diamonds? The differential equation \eqref{3.16C} is slightly unwieldy, because of the more complicated metric in finite diamonds. However, one can view a finite diamond as a subregion of an asymptotic diamond. Consider in \emph{asymptotic} diamond coordinates \eqref{3.25metric} the solution
\begin{equation}
    \Phi_\text{sourced}=\frac{E}{2}\frac{\sinh(x_\text{bdy}-\abs{x})}{\cosh(x_\text{bdy})\cosh(s)}\,,\label{4.14sol}
\end{equation}
with 
\begin{equation}
    \cosh(x_\text{bdy})=\frac{1}{\cos(\sigma_\text{bdy})}\,.
\end{equation}
According to equation \eqref{3.4B}, this determines the location of $(x_\text{bdy},s=0)$ of the boundary of a smaller causal diamond \emph{inside} an asymptotic diamond. We then see that the solution \eqref{4.14sol} satisfies both the jump conditions, and $\Phi_\text{sourced}\rvert_\text{bdy}=0$. Hence there is indeed a unique solution! Stationary (geodesically infalling) matter particles starting at locations $x_0\neq 0$ can be constructed similarly by patching together vacuum solutions with jump conditions at $x=x_0$ arising due to a source term $-E \delta(x-x_0)$ in equation \eqref{4.10source}. Note that fixed $x=x_0$ are indeed infalling geodesics in asymptotic diamonds. For some general stationary matter profile $E_\text{matter}(x)$ one then finds a convolution integral with the solutions sourced at $x$. For conformal matter the equations become more cluttered (roughly, because conformal matter is not stationary). However, we believe we have sufficiently illustrated the point. In conclusion: the space of classical dilaton solutions is \emph{not} affected by matter sources. Therefore, the gravitational and matter Hilbert spaces (prior to imposing the constraint) indeed decouple.

In particular, states with fixed $B$ (using fixed $\Phi_\text{bdy}$ boundary conditions) have a dilaton profile at the geodesic slice completely fixed by equation \eqref{3.12phi}, and the sourced contributions computed above. At fixed $B$, matter does not backreact on the dilaton profile! For asymptotic diamonds this conclusion was well-known \cite{Penington:2023dql}.

\subsection{Type II gravitational entropy}\label{sect:4.3entropy}
In the remainder of this section we investigate the gravitational interpretation of the entropy \eqref{4.7S}. We follow closely the discussion below equation \eqref{2.57entropyformula} in the case of closed universes. Combining equations \eqref{3.14k} and \eqref{3.16C} one finds:
\begin{equation}
    K_\text{grav}=\int_{-\infty}^{+\infty}\d x\,\sqrt{h}n^\mu\xi^\nu T_{\text{grav}\,\mu\nu}=-\int_\text{geo}\d x\,\sqrt{h}n^\mu\xi_\mu \Phi_\text{sourced}(x)\,.
\end{equation}
Here we used the fact that the sourced solutions $\Phi_\text{sourced}(x)$ exponentially decay towards $x=\pm \infty$, both for finite and asymptotic diamonds. See for instance equation \eqref{4.13phisource}. Using equation \eqref{4.7S}, one obtains the following expression for the gravitational entropy of our subregions:
\begin{equation}
    \boxed{\delta S=\frac{2\pi}{\hbar}\int_\text{geo}\d x\,\sqrt{h}n^\mu\xi_\mu \,\delta \Phi(x)\,}\label{4.17s}
\end{equation} 
We can be explicit about the integral using conformal coordinates $\d s^2=C^{-2}(-\d s^2+\d x^2)$. The induced metric is $\sqrt{h}=C^{-1}$, therefore the unit normal is $n^{\mu}=(C,0)$ and the CKV is $\xi^\mu = (1,0)$. The measure becomes the conformal factor of the metric:
\begin{align}
    \sqrt{h}n_\mu\xi^\mu = \frac{1}{C^2}\,.
\end{align}
For symmetric finite diamond, this factor appears in (\ref{3.6metric}). For asymmetric diamond, one uses equation (\ref{eq:asymmetricdiamond}). Here $(\sigma_\text{L},\sigma_\text{R})$ should be interpreted as functions of $(B,\Phi_\text{L},\Phi_\text{R})$ which are constants in this setup (we have indeed worked in fixed $B$ sectors in this section). For ``asymptotic'' diamonds $\sqrt{h}n^\mu\xi_\mu\rvert_\text{geo} =1$ and equation \eqref{4.17s} reduces to the closed universe equation \eqref{2.57entropyformula}, but with the spatial integration over $-\infty<x<+\infty$ now. 

We stress that this expression does \emph{not} reduce to a boundary quantity, contrary to common intuition and following the closed universe case. One might alternatively wonder if $\delta S$ is related with the changes in the value of the dilaton at the extremal point: $\delta \Phi_\text{extr}$. One computes $\delta \Phi_\text{extr}$ for instance for asymptotic diamonds by finding the minimum of the sum of \eqref{4.13phisource} and \eqref{3.25metric} and checks that the changes in entropy and in extremal area are \emph{not} proportional $\delta S \not\sim \delta \Phi_\text{extr}$.

We may contrast this situation with the calculation of gravitational entropy of a Rindler wedge \cite{Chandrasekaran:2022eqq}. Consider the JT Rindler wedge's metric and dilaton solutions:
\begin{equation}\label{eq:rindler}
    \d s^2=\d \rho^2-\sinh^2(\rho)\d t^2\,,\quad \Phi_\text{bare}=\Phi_h\cosh(\rho)\,.
\end{equation}
One of the boundaries is located at the extremal surface, where (by definition) the spatial derivative of the dilaton vanishes:
\begin{equation}
    \Phi_\rho\rvert_\text{extr}=0\,.\label{4.20ext}
\end{equation}
The other boundary is subject to usual asymptotically AdS$_2$ boundary conditions \cite{Maldacena:2016upp}. We fix $\d s=\i \d t/\varepsilon$. This time coordinate $t$ is related with the conventional JT Rindler time by $t=\Phi_h T$ and, according to equation \eqref{3.20omega}, is conjugate to $\Phi_h$. We will see this property also emerge from the following calculation. The Einstein equation \eqref{2.26eeq} in this metric describing the generator of $t$ evolution reads:
\begin{equation}\label{eq:4.21}
    \sqrt{h} n^\mu \xi^\nu T_{\mu \nu}=-\sinh(\rho)\Phi_{\rho\rho}+\sinh(\rho)\Phi
\end{equation}
Crucially, this is a total derivative, so that the associated gravitational constraint is a boundary term! This happens since $t$ flow is an actual isometry, unlike in our case where $s$ flow is a conformal isometry but not an actual isometry. Using $\Phi_\rho\rvert_{\text{extr}}=0$ and integrating over space one obtains the familiar result:
\begin{equation}
    H_\text{matter}=\Phi_\text{extr}-\frac{\Phi\rvert_\text{bdy}-\Phi_\rho\rvert_\text{bdy}}{\varepsilon}=\Phi_\text{extr}-E_\text{ADM}\,.\label{4.19H}
\end{equation}
Here one recognizes equation (A.7) in \cite{Stanford:2020wkf} for the ADM energy $E_\text{ADM}$. As $H_\text{grav}$ generates $t$ evolution, one could have anticipated finding $\Phi_\text{extr}$ (the Legendre conjugate of $t$) in this expression. As check for equation \eqref{4.19H}, note that the right-hand side vanishes for the vacuum solution $\Phi=\Phi_h\cosh(\rho)$. In the scenario where we consider configurations with $E_\text{ADM}$ fixed the gravitational entropy formula becomes (the calculation is morally similar to the one presented in section \ref{sect:2.4closedentropy})
\begin{equation}
    \delta S=\frac{2\pi}{\hbar}\delta \Phi_\text{extr}\,.
\end{equation}
This indeed reproduces $\delta S=\delta A_\text{extr}/4 G$ in JT gravity, using $\hbar\to 8\pi G$ as motivated in equation \eqref{2.70G}.

This is an example of a more general phenomena: the generators of isometries reduced to boundary terms. For conformal isometries, $\xi$ does not satisfy the required property. See Appendix \ref{sec:appxb}. Therefore the gravitational entropy (which in general is related to changes in the generator of the diffeomorphism we are dressing to) does not need to reduce to boundary terms. We are not aware of any mathematical or physical issues with dressing to conformal time. So, we conclude that gravitational entropy need not always be a boundary term!

\section{Concluding questions and speculations}\label{sect:concl}
We conclude by listing some - in our opinion - interesting open questions that originate from this work. We speculate on their resolution.

\begin{enumerate}
    \item \textbf{Algebraic entropy of closed universes versus no-boundary gravitational path integral?}

    For the current discussion we omit the quantum number $B$ in the Hilbert space. One may organize states in $\mathcal{H}_\text{grav}\otimes \mathcal{H}_\text{matter}$ by diagonalizing the ``constraint quantum number'':
    \begin{equation}
        \mathbf{C}=\mathbf{K}_\text{grav}+K_\text{matter}
    \end{equation}
    All operators $a$ in our algebra of physical operators (by definition) commute with $\mathbf{C}$. Therefore they decompose as:
    \begin{equation}
        [a,\mathbf{C}]=0\to \mathbf{a}=\int_\otimes \d c\,a_c
    \end{equation}
    These act on the similarly block-diagonal decomposition of the extended Hilbert space:
    \begin{equation}
        \mathcal{H}_\text{grav}\otimes \mathcal{H}_\text{matter}=\int_\otimes \d c\,\mathcal{H}_c\,.\label{5.3hilbert}
    \end{equation}
    Thus we have superselection sectors labeled by the eigenvalues $c$ of $\mathbf{C}$. When the matter algebra is type III and the matter state is KMS with inverse temperature $2\pi$, we may schematically write the tracial state as in equation \eqref{2.43trace}:
    \begin{equation}
        \rho_\text{NB}=e^{-2\pi \mathbf{C}/\hbar}\,.
    \end{equation}
    For operators that do not change $K_\text{grav}$ (this includes most of the interesting observables, such as those discussed in equation \eqref{2.542point}) the more refined version of the trace \eqref{2.44trace} may be written as
    \begin{equation}
        \tr(a)=\int_{-\infty}^{+\infty}\d Ke\,^{-2\pi K/\hbar}\Tr_\text{matter}(\rho_\text{thermal}\, a)=\Tr(\rho_\text{NB}\, a) =  \int_{-\infty}^{+\infty}\d c\,e^{-2\pi c/\hbar}\,\Tr_c(a_c)\,.
    \end{equation}
    In the intermediate (formal) step we leveraged the two Hilbert space decompositions \eqref{5.3hilbert}. There are two remarks we want to make about this equation.

    Firstly, this equation looks strikingly similar to equation (4.9) in \cite{Soni:2023fke} (one may note the similarity with the superselection sector structure that exists for theories with edge modes \cite{Klinger:2023tgi,Donnelly:2014fua,Blommaert:2018rsf}). In \cite{Soni:2023fke}, the author (morally) associates a nontrivial entropy with a direct product of type I algebras. In the context of the present work, this raises the question whether one could associate a gravitational entropy with matter coupled to quantum JT gravity for closed universes - even in the case where the matter is type I.\footnote{Perhaps the better analogy with \cite{Soni:2023fke} is obtained by considering the isomorphic algebra (see for instance the discussion below equation (2.6) in \cite{Chandrasekaran:2022cip}) spanned by
    \begin{equation}
        a(s)\,,\quad f(\mathbf{K}_\text{grav}-K_\text{matter})\,.
    \end{equation}
    In this conjugated algebra the constraint effectively becomes $\textbf{K}_\text{grav}$.} The gravitational path integral predicts seemingly a nontrivial gravitational entropy in some cases where the Cauchy slice is a whole closed universe, and where the matter is type I \cite{Blommaert:2025bgd}. Is there a connection?

    Secondly, naively, the gravitational path integral produces a projection operator $\delta(\mathbf{C})$ \cite{Blommaert:2025bgd}. How is this consistent with the algebraic tracial state $e^{-2\pi \mathbf{C}/\hbar}$? As discussed in figure \ref{fig:braket}, we could have chosen a different thermal matter state for our preparation, for instance, by considering a bra-ket wormhole with contour $\mathcal{C}_{n\pi}$ \cite{Chen:2020tes}. This would lead to the tracial state
    \begin{equation}
        \mathcal{C}_{n\pi}\to e^{-n\pi \mathbf{C}/\hbar}\,.
    \end{equation}
    This is not in contradiction with the uniqueness of the trace. Indeed, these matter states are not in the same Hilbert space as no finite amount of perturbations can bring one thermal state to the other (very loosely speaking). Furthermore the gravitational entropy would become
    \begin{equation}
        \mathcal{C}_{n\pi}\to\delta S=\frac{n\pi}{\hbar}\oint_\text{geo}\d x\,\delta \Phi(x)\,.
    \end{equation}
    More in general, one might consider linear combinations. Naively, the gravitational path integral instead prepares the state:
    \begin{equation}
        \mathcal{C}_\text{grav}\to\delta(\mathbf{C})=\int_{-\infty}^{+\infty}\d N\,e^{\i N \mathbf{C}/\hbar}\,.
    \end{equation}
    Is this consistent with one of the bra-ket wormhole states? Should one perhaps sum over bra-ket wormhole saddles? Does the integration over the lapse function $N$ in JT gravity localize to those saddles somehow? This seems a simple setup to investigate the issues raised in section 6 of \cite{Blommaert:2025bgd}.

     \item \textbf{Gravitational entropy for observers in closed universes with restricted causal access?} 
    
    Consider FLRW cosmologies where an observer has causal access to more than half of a Cauchy slice of the FLRW cosmology. Could one construct a type II algebra of observables, and compute the associated gravitational entropy? In section 5 of \cite{Blommaert:2025bgd}, it was emphasized that naively matter correlators in such a setup can not satisfy KMS, which makes constructing a tracial state more complicated (if even possible). This issue could potentially be sidestepped by considering rescaled conformal matter operators as in equation \eqref{2.11rescale}, following Casini-Huerta-Meyers \cite{Casini:2011kv}. Does this lead to a type II algebra with an interesting gravitational entropy, or did we throw out the baby with the bathwater by considering de-facto time-translation invariant observables? More generally it would be desirable to have any example of a type II algebra based on \emph{actually} time dependent physics.\footnote{Two examples involving time-dependent backgrounds are \cite{Kudler-Flam:2023qfl,Kudler-Flam:2024psh} but the resulting algebras again give time translation invariant correlators.} Modular flow always generates an isometry. Perhaps the crux is to decompose modular flowed operators into physically relevant simple operators.

    \item \textbf{Interior observables?}

    What happens to the diamonds studied in section \ref{sect:3.2quantumdiamonds} for $\Phi_\text{bdy}<\Phi_h$? The diamond now lie in the black hole interior. See figure \ref{fig:interior}. The phase space description is related to the exterior diamonds considered in equation \eqref{3.5B} by $\Phi_h T=\i \pi/2+\Phi_h T_\text{interior}$:
    \begin{equation}
        \sinh(B/2)=\sqrt{1-\Phi_\text{bdy}^2/\Phi_h^2}\sinh(\Phi_h T_\text{interior})\,.\label{5.10B}
    \end{equation}
    This follows from the equation \eqref{3.4B} for $B$, combined with the modified version of equation \eqref{3.2coord}:
    \begin{equation}
        \coth(\Phi_h T_\text{interior})=\frac{\sin(\tau)}{\sin(\sigma)}\,.
    \end{equation}
    The Hamiltonian $E_\text{interior}$ conjugate to $T_\text{interior}$ remains \eqref{3.22de}, but we are describing the solutions with energy above the potential barrier because $\Phi_h^2>\Phi_\text{bdy}^2$. Indeed, the classical solutions \eqref{5.10B} go through $B=0$. A particularly natural class of interior diamonds are obtained upon considering $\Phi_\text{bdy}=0$. The phase space simplifies to $E_\text{interior}=P^2$, $B=2\Phi_h T_\text{interior}$. Finite cutoff JT gravity may thus allow us to study quantum gravity in a black hole interior, and construct gauge-invariant observables in the interior. What can be learned from this? One could study wormhole effects in this context, along the lines of \cite{Ahmedwip}.
    
 \begin{figure}
 \centering
 \begin{tikzpicture}[baseline={([yshift=-.5ex]current bounding box.center)}, scale=0.7]
 \pgftext{\includegraphics[scale=1]{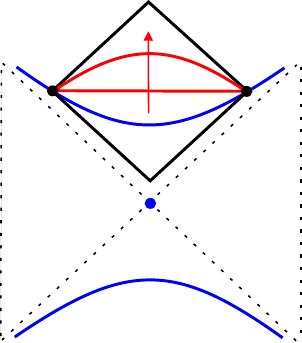}} at (0,0);
    \draw (-0,3.4) node {diamond};
    \draw (-4.3,0.4) node {asymptotic};
    \draw (-4.3,-0.3) node {boundary};
    \draw (0,-2.4) node {\color{blue}$\Phi_\text{bdy}$};
  \end{tikzpicture}
 \caption{For $\Phi_\text{bdy}<\Phi_h$ the subregions we have been describing are associated with diamonds which cover part of the black hole interior. What can an algebra of gauge-invariant interior observables teach us?}
 \label{fig:interior}
\end{figure}
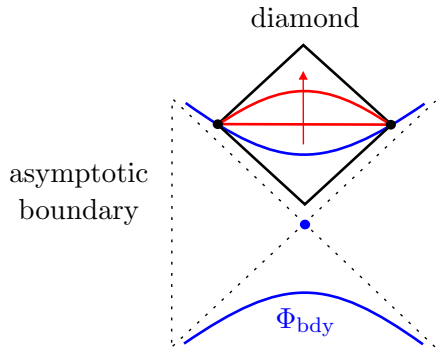

    \item \textbf{Probing interior singularities using dilaton clocks?}
    
    Consider black hole interiors in some general 2d dilaton gravity \cite{Witten:2020ert,Witten:2020wvy} with potential $V(\Phi)$. The interior metric is conveniently expressed using the dilaton as a time coordinate:
    \begin{equation}
        \d s^2=-\frac{\d \Phi^2}{(F(\Phi_h)-F(\Phi))}+(F(\Phi_h)-F(\Phi))\,\d T_\text{interior}^2\,.
    \end{equation}
    Here $F'(\Phi)=V(\Phi)$. We consider monotonically increasing functions $F(\Phi)$ that describe a dilaton gravity solution with a curvature singularity at some finite dilaton value $\Phi_\text{sing}$:
    \begin{equation}
        F(\Phi_\text{sing})=-\infty\quad \to \quad  R(\Phi_\text{sing})=+\infty\,.
    \end{equation}
    The interior in these cases is the entire coordinate patch
    \begin{equation}
        \Phi_\text{sing}<\Phi<\Phi_h\,.
    \end{equation}
    In the main text, we used the extrinsic curvature $K$ of the spacetime as a gauge-invariant clock. We now explain that, alternatively, for closed universes and for black hole interiors, one could use the dilaton $\Phi$ as a physical clock. Adhering to the notation of section \ref{sect:2obcl} we denote this clock by $S$
    \begin{equation}
        \Phi=S\,.
    \end{equation}
    We consider a finite interval for $T_\text{interior}$. In practice, we may make the spatial coordinate periodic $T_\text{interior}\sim T_\text{interior}+\beta$. Using $N=1/\sqrt{F(\Phi_h)-F(\Phi)}$) one finds the minisuperspace variables
    \begin{equation}
        \ell=\beta \sqrt{F(\Phi_h)-F(\Phi)}\,,\quad k=-\beta \frac{V(\Phi)}{2}\,,\quad p=\sqrt{F(\Phi_h)-F(\Phi)}\,.
    \end{equation}
    The satisfy the gravitational constraint \eqref{2.20hgrav}:
    \begin{equation}
        H_\text{grav}=-p k -\ell \frac{V(\Phi)}{2}=0\,.
    \end{equation}
    This may be rewritten acting on wavefunctions $\psi(\Phi,p)$ as
    \begin{equation}
        \frac{\d}{\d \Phi}-\frac{V(\Phi)}{2 p}\frac{\d}{\d p}=0\,\label{5.18}.
    \end{equation}
    Introducing the change of variables
    \begin{equation}
        \Phi=S\,,\quad p=\sqrt{F(\Phi_h)-F(S)}\,,
    \end{equation}
    This constraint acting on wavefunctions $\psi(S,\Phi_h)$ becomes
    \begin{equation}
        \frac{\d}{\d S}=0\,.
    \end{equation}
    This implies that the generator of $S\to S+c$ (changes in dilaton) is a constraint, hence the dilaton $\Phi=S$ may be used as a gauge-invariant clock to which to relate observables. For general dilaton gravity, indeed, dressing to curvature seems rather complicated because the potential term in the gravitational constraint is non-linear in $\Phi$ (the conjugate of curvature). On the constraint phase space the symplectic form becomes
    \begin{equation}
        \omega_\text{physical}=\d \frac{V(\Phi_h)\beta}{2}\wedge \d \Phi_h\,.
    \end{equation}
    Indeed, the combination $V(\Phi_h)\beta/2$ is naturally some boost angle conjugate to $\Phi_h$.

    In general dilaton gravity models, spacetime curvature depends on this physical time coordinate:
    \begin{equation}
        R(S)=-V'(S)\,.
    \end{equation}
    Therefore we are essentially placing operators at well-defined spacetime curvature. Could one use the dilaton clock to construct an algebra of gauge invariant observables in such a curved interior? What happens when operators approach the curvature singularity? Is the effect of the singularity resolved by time-smearing due to gravitational dressing, as in equation \eqref{2.55smeared}?

    Relatedly, one could attempt to understand how the Hilbert space and the algebra of observables is affected by including effects of topology change in the construction of the gravitational Hilbert space \cite{Iliesiu:2024cnh}, including now also the $S$ variable.

    \item \textbf{Generalized second law from asymmetric diamonds?} 
    
    We proposed an equation \eqref{4.17s} for the entropy of a subregion in JT gravity coupled to conformal field theory. A natural question is whether this entropy follws the generalized second law \cite{bekenstein1973black} (or GSL). One could for instance investigate whether the second law holds for the diamonds sketched in figure \ref{fig:gsl}. Here we imagine $\Phi_\text{R}\to +\infty$ and keeping $\Phi_\text{L}$ fixed (one could also fix the left endpoint to some extremal surface as in equation \eqref{4.20ext}). The size $B$ of the subregion decreases with $\tau_\text{bdy}$. How does the entropy change? It is not immediately clear to us which dilaton solutions to consider and compare between diamonds with different $\tau_\text{bdy}$. Therefore we leave this as an open question. A naive attempt, fixing $\Phi_\rho\rvert_\text{L}=0$ and keeping $\delta \Phi$ unchanged appears to result in a monotonically \emph{decreasing} entropy. This should likely not be taken seriously.

    \begin{figure}
        \centering
        \begin{tikzpicture}[baseline={([yshift=-.5ex]current bounding box.center)}, scale=0.7]
 \pgftext{\includegraphics[scale=1]{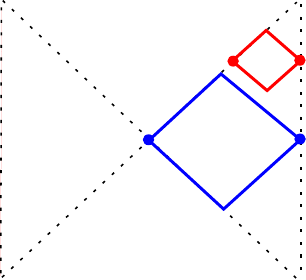}} at (0,0);
    \draw (3.8,0) node {\color{blue}$\tau_\text{bdy}=0$};
    \draw (3.8,1.3) node {\color{red}$\tau_\text{bdy}>0$};
  \end{tikzpicture}
        \caption{One version of the generalized second law would be if $S_\text{red}\geq S_\text{blue}$. Is this correct?}
        \label{fig:gsl}
    \end{figure}

    \item \textbf{Alternative subregion definitions?}

    We have defined subregions in quantum gravity by fixing $\Phi\rvert_\text{bdy}=\Phi_\text{bdy}$. This is (morally) similar to considering fixed area states in quantum gravity \cite{Dong:2018seb,Dong:2022ilf}. One could wonder if other reasonable definitions of subregions exist, and could be worked out. One seemingly reasonable example would be to fix the momentum $p$ conjugate to boundary proper length:
    \begin{equation}
        p^2=g^{\mu\nu}\nabla_\mu \Phi\nabla_\nu \Phi=\Phi^2-\Phi_h^2
    \end{equation}
    One can imagine fixing $p\rvert_\text{bdy}=p_\text{bdy}$. Curves of fixed $P_\text{bdy}$ are identical to curves of fixed dilaton. The phase space description changes a bit due to explicit $\Phi_h$ dependence in $p_\text{bdy}=\sqrt{\Phi_\text{bdy}^2-\Phi_h^2}$. This affects the Schrodinger equation for states with fixed Rindler energy \eqref{3.21}. However, naively, our derivation of the entropy formula \eqref{4.17s} remains valid. It would be interesting to investigate this and other alternative definitions further. Fixing $P_\text{bdy}$ has as advantage that the usual extremal surface description becomes simply the boundary condition
    \begin{equation}
        p_\text{bdy}=0\,.
    \end{equation}
    We thank Arvin Shahbazi-Moghaddam for discussions on this point.

    \item \textbf{Towards static patch quantum mechanics?}

    This last item is the most speculative and least worked out. Could one attempt to discuss quantum mechanics for diamonds associated with subregions in a dS static patch ending on regions of fixed dilaton? The solutions are
    \begin{equation}
        \d s^2=\frac{\d \sigma^2-\d \tau^2}{\cosh(\sigma)^2}\,,\quad \Phi=\Phi_h\tanh(\sigma)\,.
    \end{equation}
    We consider a subregion starting at the pode $\Phi=0$ and ending at $\Phi_\text{bdy}$. To have some notion of quantum mechanics, one might want to start with finding the length of a geodesic describing this region that changes as function of some time coordinate $B(\Phi_h,T)$. Static patch time $\tau$ would not work (because this is an isometry). However, one can maybe consider global proper time $T$. How does a geodesic length of a subregion ending at fixed $T$ on the pode and on a $\Phi_\text{bdy}$ surface evolve as function of $T$? What is the conjugate to $T$ (two boundaries contribute naively). What are the resulting wavefunctions?
    
\end{enumerate}

\section*{Acknowledgments}
We thank Ven Chandrasekaran, Julian De Vuyst, Jonah Kudler-Flam, Adam Levine, Thomas Mertens, Jacopo Papalini, Geoff Penington, Arvin Shahbazi-Moghaddam, Bruno Torres and Aron Wall for discussions. AB was supported by the Leinweber foundation, the US DOE DE-SC0009988, and the Sivian fund. CHC was supported by AFOSR award FA9550-22-1-0098.

\appendix

\section{Only isometries guarantee gravitational boundary charges}\label{sec:appxb}
A general lore is that gravitational constraints lead to boundary charges. As stated at the end of section \ref{sect:4obssubr}, we now show that this is only guaranteed if the vector field generating the diffeomorphism associated with the constraint is a Killing vector. Equivalently, only diffeomorphisms that represent isomorphisms result in gravitational charges located on the boundary (or edge) of the (sub)region of interest. General diffeomorphisms result in bulk charges! Examples are conformal time translations in the diamonds of section \ref{sect:3diamons} which are generated by a conformal Killing vector. Not a Killing vector! Hence, the charge is not a boundary quantity, and neither is the associated type II entropy.

We define the gravitational charge associated with a vector field $\xi^\mu$ to be the Einstein's tensor $G_{\mu\nu}$, contracted with the vector field $\xi^\mu$ and the normal of a co-dimension $1$ Cauchy slice $n^\nu$ integrated along the slice. In JT gravity with matter, we called the equivalent of Einstein's tensor \eqref{2.26eeq}:
\begin{align}\label{eq:b1}
    T_{\text{grav }\mu\nu} = \nabla_\mu\nabla_\nu\Phi -g_{\mu\nu}\nabla^2\Phi+g_{\mu\nu}\Phi.
\end{align}
We may then define the gravitational current:
\begin{align}
    j^\mu_{\rm grav}= T_{\text{grav }}^{\mu\nu} \xi_\nu\,.
\end{align} 
We will show that this current equals a total divergence plus a purely bulk term:
\begin{align}\label{eq:jmu}
    j_{\rm grav}^\mu= \nabla_\nu \, k^{\mu\nu} + \frac{1}{2}\bigg( X^{\mu\nu}\nabla_\nu\Phi-\Phi\nabla_\nu X^{\mu\nu} \bigg)\,.
\end{align}
Here
\begin{align}
    k^{\mu\nu} = \xi^\nu\nabla^\mu \Phi-\xi^\mu\nabla^\nu\Phi - \Phi \nabla^{[\mu}\xi^{\nu]}\,,\quad X^{\mu\nu} = h^{\mu\nu} - h g^{\mu\nu},\quad h_{\mu\nu}=\mathcal{L}_\xi g=2\nabla_{(\mu}\xi_{\nu)}\,.\label{b.4k}
\end{align}
Integrating the current (\ref{eq:jmu}) over the slice, we get the gravitational constraint (or charge):
\begin{align}
\boxed{
    K_{\rm grav} = Q_\xi|_{\rm bdy} + \int \d x\sqrt{h}\,n_\mu\frac{1}{2}\bigg( X^{\mu\nu}\nabla_\nu\Phi-\Phi\nabla_\nu X^{\mu\nu} \bigg)\,}\label{b.5charge}
\end{align}
Here\footnote{Here $\epsilon_{\mu\nu}$ is the Levi-Cevita symbol, which contains some factor of the square root determinant of the metric $\epsilon_{12} = \sqrt{-g}$.}
\begin{align}\label{eq:qxi}
    Q_\xi = -\frac{1}{2}\epsilon_{\mu\nu}k^{\mu\nu}.
\end{align} 
Before proving equation \eqref{b.5charge} we discuss its interpretation. The main point is that if $\xi$ is Killing, then the bulk term vanishes. Indeed, for $\xi$ generating some isometry we have:
\begin{equation}
    h_{\mu\nu\,\text{iso}}=2\nabla_{(\mu}\xi_{\nu)}=0\,.
\end{equation}
Therefore isometries \emph{do} guarantee that the associated gravitational constraint is a boundary charge:
\begin{equation}
    \boxed{K_{\text{grav}\,\text{iso}} = Q_\xi|_{\rm bdy}\,}
\end{equation}
Otherwise, the gravitational constraint is generally \emph{not} a boundary charge!\footnote{Strictly speaking, the bulk term vanishes whenever $\frac{X^{\mu\nu}}{\Phi}=\mathrm{const},$ however we did not obtain interesting solutions besides the isometries} The requirement is that the vector field is Killing on the slice where the charge is defined, e.g., a Cauchy slice or event horizon. Notice that the vector field does not have to be Killing everywhere but just on the slice\footnote{An example is \cite{Klinger:2026tws}. They include modes associated with "supertranslations" on the horizon, which are not isometries of the full spacetime. However, they are Killing on the horizon. Because they only needed the constraint integrated along the horizon, our formula confirms that the gravitational modes can be treated as boundary terms, resulting in the ``ordinary'' form of generalized entropy as a boundary quantity. This is not the general situation, however.}.

A similar situation arises for Gauss' EM constraint. The constraint is $\nabla\cdot E = \rho$ with $E$ the electric field and $\rho$ the electric charge density. We may integrate the constraint against a function $\lambda(x)$ on some slice $\mathcal C$:
\begin{align}
    \int_\mathcal{C} \d x\,\lambda(x)\,(\nabla\cdot E - \rho )=0
\end{align}
Partial integration results in the analogue of equation \eqref{b.5charge}:
\begin{align}
    \oint_{\partial\mathcal{C}} \d u\,\lambda(u)\,E_\perp- \int_\mathcal{C} \d x\,\nabla\lambda(x)\cdot E = \int_\mathcal{C} \d x\,\lambda(x)\rho
\end{align}
It is clear if $\nabla \lambda$ is orthogonal to the electric field $E(x)$ on the slice $x\in\mathcal{C},$ then the bulk term vanishes, and we have a Gauss-like constraint. But the most general matter charge component is \emph{not} determined solely by the boundary configuration of the perpendicular electric field $E_\perp$.

We now prove \eqref{b.5charge}. To get \eqref{b.5charge} from \eqref{eq:jmu} we observe that $k^{\mu\nu}$ is anti-symmetric and that in two dimensions, any anti-symmetric rank-$2$ tensor is proportional to the Levi-Cevita symbol. The prefactor is
\begin{align}
    k^{\mu\nu} = Q_\xi \epsilon^{\mu\nu}\,,
\end{align}
which can be checked easily by plugging this into (\ref{eq:qxi}) and using $\epsilon_{\mu\nu}\epsilon^{\mu\nu}=-2.$ One then indeed obtains
\begin{align}
    \int \d x\sqrt{h}\,n_\mu \nabla_\nu k^{\mu\nu}= \int \d\ell \,n_\mu\,\epsilon^{\mu\nu}\nabla_\nu Q_\xi= \int \d\ell\, \ell^\nu\nabla_\nu Q_\xi=\int \d\ell\, \frac{\d}{\d \ell} Q_\xi=Q_\xi|_\text{bdy}\,.
\end{align} 
In the first equality we used that the covariant derivatives of Levi-Cevita symbol vanish and introduced the proper length $\sqrt{h}\,\d x=\d \ell$. In the second equality we introduced the unit tangent vector $\ell^\nu =\epsilon^{\mu\nu}n_\mu$. This leaves us with proving \eqref{eq:jmu}, which we do by brute force expansion. Firstly:
\begin{align}
    \nabla_\nu k^{\mu\nu} + \frac{1}{2}X^{\mu\nu}\nabla_\nu\Phi = \xi^\nu\nabla^\mu\nabla_\nu\Phi - \xi^\mu\nabla^2\Phi-\Phi\nabla_\nu\nabla^{[\mu}\xi^{\nu]}.
\end{align} The first two terms on the right match the first two terms in (\ref{eq:b1}). To show that the third term cancels with the third term in (\ref{eq:jmu}), we use the definition of curvature as a commutator: $R_{\mu\nu}\,\xi^\nu = [\nabla_\mu,\nabla_\nu]\,\xi^\nu$ to rewrite
\begin{align}
    \nabla_\nu\nabla^{[\mu}\xi^{\nu]} = \frac{1}{2}\bigg( \nabla^\mu\big(\nabla\cdot\xi\big)+R^\mu{}_\rho\,\xi^\rho-\nabla^2\xi^\mu  \bigg)\,,\quad \nabla_\nu X^{\mu\nu} = -\nabla^\mu\big(\nabla\cdot\xi\big)+R^\mu{}_\rho\xi^\rho + \nabla^2\xi^\mu\,.
\end{align} 
The last ingredient that we need is that for AdS$_2$ the curvature is simply $R^\mu {}_\rho = - \delta^\mu{}_\rho$, so the curvature term becomes $\xi^\mu\Phi$, which reproduces the last term in (\ref{eq:b1}). This concludes our derivation.

\subsection{Examples}

As check on our formula \eqref{b.5charge} we work out some examples. First, let's do the Rindler wedge. We recall from (\ref{eq:rindler}) that the metric is $\d s^2=\d\rho^2-\sinh(\rho)^2 \d t^2.$ A simple calculation of Christoffel symbols leads to
\begin{align}
    \nabla^\rho\xi^t=-\nabla^t\xi^\rho=\coth(\rho)\,.
\end{align}
Inserting this into equation \eqref{b.4k} results in
\begin{align}
    k^{t\rho} = -\Phi_\rho + \coth(\rho)\Phi\,,
\end{align} which needs to be contracted with the Levi-Civita symbol, $\epsilon_{t\rho}=\sqrt{-g}=\sinh(\rho)$. The boundary charge is then in agreement with (the integral of) \eqref{eq:4.21}:
\begin{align}
    Q_\xi = \sinh(\rho) \Phi_\rho-\cosh(\rho)\Phi\,,
\end{align}

Next consider a general 2d conformal metric with conformal time $s$
\begin{align}
    \d s^2 = \frac{-\d s^2+\d x^2}{C^2},
\end{align}
We study the charge associated with $s$ translations, as throughout this paper. The Christoffel symbols are
\begin{align}
    \Gamma^x_{ss} = \Gamma^s_{xs} = -\frac{C_x}{C}\,.
\end{align}
Using $\Phi\nabla^{[s}\xi^{x]} = \Phi\cdot C\,C_x$ one finds $k^{sx} = -C^2\Phi_x-\Phi\cdot C\,C_x$. Therefore, the boundary charge becomes
\begin{align}\label{eq:qckv}
    Q_\xi = \Phi_x + \frac{C_x}{C}\Phi\,.
\end{align} 
Crucially the bulk term in \eqref{b.5charge} is nonzero! For a CKV
\begin{align}
    X^{\mu\nu}_\text{CKV} = -\big(\nabla\cdot \xi\big)g^{\mu\nu}\,.
\end{align} 
Plugging this into the conformal metric one finds $ X^{\mu\nu} = 2 C\, C_s \eta^{\mu\nu}$. Using the Christoffel symbols, one finds that the bulk contribution in $K_{\rm grav}$ becomes:
\begin{align}\label{eq:kbulk}
    K_\text{grav}\supset \int^{+\infty}_{-\infty} \d x \bigg\{ \frac{C_s}{C}\Phi_s - \Phi\bigg(\frac{C_{ss}}{C}-\frac{C_s^2}{C^2}\bigg)\bigg\}\,.
\end{align}
For asymptotic diamonds this simplifies to
\begin{align}
K_{\rm grav }= 2 \Phi_x(x\to \infty)+\int^{+\infty}_{-\infty} \d x\,\bigg\{ \tanh(s)\Phi_s-\frac{1}{\cosh(s)^2}\Phi\bigg\}\,.
\end{align} 
This further simplifies using the explicit solution \eqref{3.25metric}, which has the property $\Phi_{x\,\text{bare}}(x\to \infty)\to \Phi_\text{bdy}$. As a sanity check, one can plug the bare dilaton solution into the bulk charge and check that it cancel out with the boundary charge: $K_{\rm grav}=0.$ In the presence of matter (obeying certain fall-off conditions), the boundary charge doesn't change. For example, with a particle of energy $E$ at $x=0$, the $x$ derivative of the sourced solution (\ref{4.13phisource}) exponentially decay in $x$, and their contribution to the boundary charge is $0$. For symmetric finite diamonds, one obtains similarly:\footnote{We argue that $\Phi_x$ vanishes for $x\to \pm \infty$. Near the corners $x=\pm\infty,$ the infinitesimal proper length $d\ell$ grows exponential in the coordinate $x$, namely $\d\ell\sim e^{-|x|}\,\d x$. This implies that if $\Phi_x$ asymptotes to some non-zero constant, say $\lim_{x\to\infty}\Phi_x = L\neq0,$ then the rate of $\Phi$ in any physical coordinates, like the proper length, $\Phi_\ell \sim e^{|x|}\Phi_x$ would diverge as $\sim e^{|x|}L$ near the corners. Therefore indeed $\Phi_x$ must indeed damp exponentially such that $\Phi_\ell$ remains finite.}
\begin{align}
    Q_\xi(x) = \Phi_x + \frac{\sinh(x) \cos(\sigma_{\rm bdy})}{\cosh(s) + \cosh(x) \cos(\sigma_{\rm bdy})}\Phi\quad \to\quad  Q_\xi\rvert_\text{bdy}= \pm \Phi_\text{bdy}
\end{align}
This is the usual statement that gravitational boundary charge is boundary area. One finds
\begin{align}
\boxed{K_{\rm grav}\rvert_\text{geo}= 2\Phi_{\rm bdy}-\int_{-\infty}^{+\infty}\d x\,\frac{\Phi(0,x)}{1+\cosh(x)\cos(\sigma_\text{bdy})}\,}
\end{align} 
As sanity check on this (complicated) bulk contribution one may plug in the vacuum dilaton solution, which we checked results indeed in $K_{\rm grav}=0$.

In summary, whilst boundary charges measure boundary area, general gravitational constraints not associated with isometries involve non-zero bulk contributions too.

\section{Finiteness of smeared correlators near singularities}\label{app:finitecorr}
Consider time smeared conformal matter correlators on the Lorentzian closed universe \eqref{2.6metric} of the type discussed in equation \eqref{2.55smeared}:
\begin{equation}
\int_{-\infty}^{+\infty}\d S\,\abs{\phi(S)}^2\,G_{\text{torus}\,\Delta}(s_1-s_2+S\rvert \tau(B))\,.
\end{equation}
The goal of this appendix is to show that these time smeared correlators are generally \emph{finite}. Without spatial identifications of the spacetime, this is a somewhat standard property \cite{Witten:2023xze}. However, one could fear that the sum over images causes problems, signaling the breakdown of QFT near a singularity \cite{Abdalla:2025gzn}. We now demonstrate that for our specific correlators, this breakdown does \emph{not} arise.

Imagine that in the Fourier domain, the smearing function imposes some high energy cutoff. Then finiteness of the time smeared correlator boils down to asking if the Fourier transform itself is finite for every $K$:
\begin{equation}
    G_{\text{torus}\,\Delta}(K\rvert \tau(B))=\int_{-\infty}^{+\infty}\d S\,e^{\i K S}\,G_{\text{torus}\,\Delta}(S\rvert \tau(B))\,.
\end{equation}
We consider a non-interacting CFT. This means we consider one or more massless scalars. We believe this example suffices to make our point. In this case, the torus correlator is obtained by simply summing over images:
\begin{equation}
    G_{\text{torus}\,\Delta}(S\rvert \tau(B))=\sum_{m=-\infty}^{+\infty}(2\cosh(m B)-2\cosh(S))^{-\Delta}\,.
\end{equation}
One might be worried about issues arising at high energies $K$. At high energies, the Fourier transform is dominated by the most non-analytic features of the correlator. Those occur near the poles: $S\sim m B$. Expanding close to the poles one obtains (throughout this section we ignore $K$-independent prefactors):
\begin{equation}
    G_{\text{torus}\,\Delta}(K\rvert \tau(B))\to \sum_{m\neq 0}\frac{1}{\sinh(m B)^\Delta}\int_{-\infty}^{+\infty}\d S\,e^{\i K S}\,\frac{1}{(S-mB)^\Delta}\to  K^{\Delta-1}\sum_{m\neq 0}\frac{e^{\i m B K}}{\sinh(m B)^\Delta}\,.\label{Asum}
\end{equation}
To evaluate the integral, we took into account the $\i\varepsilon$ prescription for the correlator. The standard term $m=0$ contributes for large $K$
\begin{equation}
    \int_{-\infty}^{+\infty}\d S\,\frac{e^{\i K S}}{\sinh(S)^{2\Delta}}\to K^{2\Delta -1}\,.
\end{equation}
The question is whether or not the sum over $m\neq 0$ leads to divergent contributions. They do not. We could simulate the worst case scenario by replacing the sum over $m$ by an integral. Then one concludes
\begin{equation}
    G_{\text{torus}\,\Delta}(K\rvert \tau(B))_{m\neq 0}\to < K^{2\Delta -1}
\end{equation}
Since this behavior is already present in the $m=0$ contribution, we conclude this is the dominant large $K$ behavior for the correlator. The large $K$ behavior is therefore \emph{not} affected (let alone dominated) by the sum over lightcone divergences (which arise due to the spatial periodicity of the spacetime).

Note that convergence of the sum in \eqref{Asum} would be less obvious if one would have studied ordinary conformal matter operators, instead of the Casini-Huerta-Meyers rescaled operators \eqref{2.11rescale}. This would lead to extra factors $\cosh(s_1(S=mB))^\Delta \cosh(s_2(S=mB))^\Delta$ in the summation over $m$ which eliminates the exponential damping. Yet, naively, large $K$ is dominated by small values of $m$, thus the conclusion would appear to be unchanged.

The lack of high energy divergences associated with the Milne singularity may rely on the fact that we have smeared in conformal time. In proper time $T$, lightcone singularities accumulate and thus give rise to extra high-energy Fourier components. It would be interesting to understand whether smearing in proper time also results in finite correlators. This is beyond our current scope.

\bibliographystyle{ourbst}
\bibliography{Refs}

\end{document}